\documentclass[prx,a4paper,aps,twocolumn,superscriptaddress,10pt]{revtex4-1}

\usepackage{amssymb,amsmath,amsthm,graphicx,xcolor,times,xfrac,booktabs,mathtools,enumitem,xr,subfigure,bbm,verbatim,appendix,placeins}
\usepackage[unicode=true,bookmarks=true,bookmarksnumbered=false,bookmarksopen=false,breaklinks=false,pdfborder={0 0 1}, backref=false,colorlinks=true]{hyperref}
\renewcommand*{\url}[1]{\href{#1}{#1}}

\makeatletter
\theoremstyle{plain}
\newtheorem{thm}{\protect\theoremname}
\theoremstyle{plain}

\theoremstyle{definition}
\newtheorem{example}{\protect\examplename}
\theoremstyle{plain}

\theoremstyle{remark}
\newtheorem*{rem*}{\protect\remarkname}
\theoremstyle{plain}

\theoremstyle{plain}
\newtheorem{cor}[thm]{\protect\corollaryname}
\theoremstyle{definition}

\theoremstyle{plain}
\newtheorem*{thm*}{\protect\theoremname}
\theoremstyle{plain}
\newtheorem*{lem*}{\protect\lemmaname}
 
\providecommand{\propositionname}{Proposition}
\providecommand{\theoremname}{Theorem}
\providecommand{\examplename}{Example}
\providecommand{\lemmaname}{Lemma}
\providecommand{\remarkname}{Remark}
\providecommand{\conjecturename}{Conjecture}
\providecommand{\definitionname}{Definition}
\providecommand{\corollaryname}{Corollary}
\allowdisplaybreaks

\def\bra#1{\langle{#1}\vert}
\def\ket#1{\vert{#1}\rangle}

\def\BraVert{\egroup\,\mid\,\bgroup}

\def\tr#1{\mbox{tr}\left[{#1}\right]}

\newcommand{\ptr}[2]{\mbox{tr}_{#1}\left[ #2 \right]}

\newcommand{\inp}{\texttt{i}}
\newcommand{\out}{\texttt{o}}

\begin{document}

\title{The Structure of Quantum Stochastic Processes with Finite Markov Order}
\date{\today}

\author{Philip Taranto}
\email{philip.taranto@monash.edu}
\affiliation{School of Physics \& Astronomy, Monash University, Victoria 3800, Australia}

\author{Simon Milz}
\affiliation{School of Physics \& Astronomy, Monash University, Victoria 3800, Australia}

\author{Felix A. Pollock}
\affiliation{School of Physics \& Astronomy, Monash University, Victoria 3800, Australia}

\author{Kavan Modi}
\affiliation{School of Physics \& Astronomy, Monash University, Victoria 3800, Australia}


\begin{abstract}
Non-Markovian quantum processes exhibit different memory effects when measured in different ways; an unambiguous characterization of memory length requires accounting for the sequence of instruments applied to probe the system dynamics. This instrument-specific notion of quantum Markov order displays stark differences to its classical counterpart. Here, we explore the structure of quantum stochastic processes with finite memory length in detail. We begin by examining a generalized collision model with memory, before framing this instance within the general theory. We detail the constraints that are placed on the underlying system-environment dynamics for a process to exhibit finite Markov order with respect to natural classes of probing instruments, including deterministic (unitary) operations and sequences of generalized quantum measurements with informationally-complete re-preparations. Lastly, we show how processes with vanishing quantum conditional mutual information form a special case of the theory. Throughout, we provide a number of representative, pedagogical examples to display the salient features of memory effects in quantum processes.
\end{abstract}
\maketitle




\section{Introduction}\label{sec:intro}

Complex processes often exhibit genuine memory effects on timescales that cannot be ignored~\cite{BreuerPetruccione,StochProc}; however, these effects are typically limited in duration for many physical processes. For classical stochastic processes, the notion of memory length can be captured formally through the concept of \textit{Markov order}, $\ell$. This dictates that the statistics describing a system of interest at a given time only depend upon knowledge of its past $\ell$ observed states. Markov order thus provides an operationally meaningful timescale for temporal correlations, the importance of which cannot be overstated, because of its significance in reducing modeling complexity: one must only estimate the conditional ``transition'' probabilities from the most recent set of observations, rather than the exponentially many more parameters for each additional timestep further back in the history. 

Attempting to generalize the notion of Markov order to quantum processes, one immediately faces the following problem: here, there is a continuous family of possible non-commuting observables that could be measured, and the choice of measurement at any point in time (or even whether to measure at all) can directly affect the future statistics~\cite{Modi2011, Modi2012, Modi2012A,Milz2017KET}. Indeed, in quantum mechanics, one must necessarily disturb the system in order to observe realizations of the process, breaking an implicit assumption in the description of classical stochastic processes. This problem has irked the open systems and quantum information communities for some time, leading to various incompatible definitions of important concepts, such as Markovianity~\cite{Chruscinski2011,Rivas2014, Breuer2016, Li2018}.

The aforementioned issue can be remedied by accounting for the multi-time statistics corresponding to all possible sequences of interrogating instruments, which track both the measurement outcome observed and the subsequent update to the state of the system. For example, the spin state of an electron evolving in an external magnetic field can be uniquely characterized by recording the probability of the spin being found in alignment with any sequence of independent directions an experimenter might choose to measure. Such accounting can be achieved within various related modern frameworks for describing general quantum processes~\cite{Chiribella2008, Chiribella2008-2, Chiribella2009, Oreshkov2012, Costa2016, Oreshkov2016, Allen2017, Ringbauer2017, Milz2017, Pollock2018A}. In short, these frameworks describe a quantum stochastic process as a collection of joint probability distributions over the outcomes of \emph{any possible sequence of measurements} by distinguishing what one has control over, \textit{i.e.}, the instruments applied to probe the system, from the uncontrollable underlying process at hand. 

This precise characterization of quantum stochastic processes leads naturally to a set of necessary and sufficient conditions for a process to be Markovian, \textit{i.e., memoryless}~\cite{Pollock2018L}. Thus equipped, one can address the concept of \emph{memory length} by unambiguously generalizing the notion of Markov order to the realm of quantum mechanics. The intuition behind quantum Markov order remains unchanged---as in the classical case, the question boils down to whether the future statistical evolution of the system can be deduced completely, in principle, from the most recent $\ell$ instruments applied. When the state of the system is independent of any previous history following the application of some sequence of instruments, the process exhibits conditional independence between the future and history. This guarantees that \emph{any} statistics one might obtain during future measurements will be independent of those measured in the history, given knowledge of the most recent $\ell$ instruments.

In an accompanying Letter~\cite{Taranto2018}, we have used the process tensor framework~\cite{Pollock2018A,Pollock2018L} to address these issues and formally extended the notion of Markov order to the quantum realm. There, we prove that demanding that the future state be independent of its history, upon application of \textit{any} possible sequence of instruments allowable in quantum mechanics, is too strong a restriction, immediately leading to a trivial theory: no non-Markovian quantum process can display finite Markov order with respect to all possible interventions. Consequently, it is natural to study processes that have non-trivial Markov order with respect to a specified sequence of instruments. To render any future state independent of its history for such processes, one must apply the correct interrogation sequence; if there exists such a sequence, we say that the process has \emph{finite-length memory} or \emph{finite Markov order} with respect to the \emph{history-blocking sequence} in question. Indeed, quantum theory permits a rich landscape of memory effects, with many properties that distinguish it from the classical setting. In this Article, we examine the structure of finite-memory quantum processes in detail. 

In particular, we ask which kinds of processes can have finite-length memory, and what can be inferred about the underlying process through knowledge of the history-blocking sequence. We begin, in Section~\ref{sec:collisionmodel}, by motivating the study of such processes through a generalized collision model and show how it displays finite Markov order with respect to a natural sequence of information-discarding instruments. We then introduce the necessary ingredients to formulate the general theory of quantum Markov order in Section~\ref{sec:framework}, before outlining the constraints on the structure of finite-memory processes that can be deduced through knowledge of the history-blocking sequence in Section~\ref{sec:qmemory}. Along the way, we introduce a variety of representative examples, many of which have uniquely quantum properties, such as history-blocking through sequences of unitary transformations or generalized (non-orthogonal) quantum measurements. In Section~\ref{sec:vanishingqcmi}, we explore the relation between processes with finite quantum Markov order and the quantum conditional mutual information, with the main result demonstrating how processes with vanishing quantum conditional mutual information are a special case within the theory of quantum Markov order. Lastly, in Section~\ref{sec:classical}, we illuminate how memory length in classical stochastic processes is also instrument-dependent when one cannot trust the resolution of their measurement device. Although this issue of fuzzy measurements obfuscating the memory length in the classical realm is, in principle, liftable, it is fundamentally unavoidable in quantum mechanics and must be acknowledged.


\section{Memory Length of a Generalized Collision Model}\label{sec:collisionmodel}


\begin{figure*}
\centering
\includegraphics[width=\linewidth]{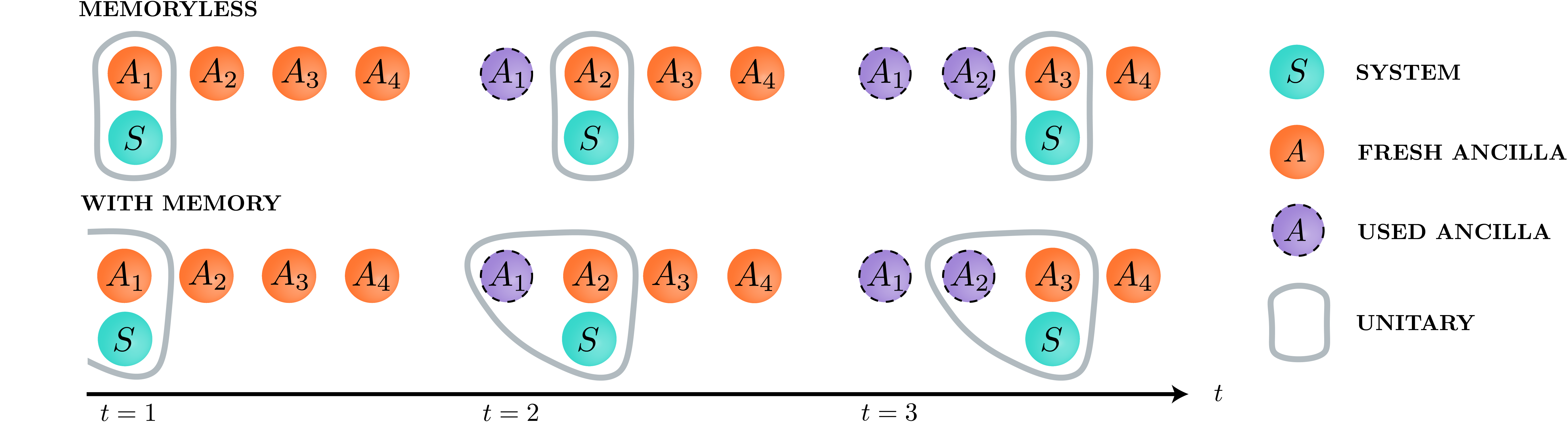}
\caption{\emph{Generalized collision model with memory.} The top row depicts a standard memoryless collision model. With time running left to right, the system $S$ (green) interacts unitarily at each timestep once with each of a number of uncorrelated, fresh ancillary states $A_j$ that constitute the environment (orange); the collision is represented by the gray boundary. Following the dynamics at $t=1$, the $A_1$ ancilla has been used and so stores information about the initial state of the system, indicated by the purple color and dashed outline (see $A_1$ at $t=2$). However, each successive portion of evolution proceeds through an interaction with a fresh ancilla that has not yet interacted with the system. Thus, any memory of the system's history cannot influence the future evolution, leading to Markovian dynamics. The bottom row shows a generalized collision model, where the system is allowed to interact with multiple ancillas between timesteps. Here, once the system interacts with $A_0$ and $A_1$ after $t=1$, again, these ancillary states can store information about the initial system state. The next step of dynamics following $t=2$ involves $A_1$ again; thus, the future dynamics are conditioned on the history. In this way, the ancillas serve to propagate memory effects through the process. \label{fig:collisionmodel}}
\end{figure*}


\subsection{Classical Markov Order}

We begin with a brief explanation of Markov order and memory length in the classical setting to lay the foundations of the concepts explored throughout this Article. Consider the toy classical process of a perturbed coin. In this example, we have a coin resting on a piece of cardboard, which is being gently shaken at discrete timesteps $k \in \{ 1, \hdots, n \}$, resulting in a time-independent probability, $p > 1-p$, for the coin to retain its previous orientation between each shake; with probability $1-p$, the coin flips from heads ($\texttt{H}$) to tails ($\texttt{T}$) or vice versa. It is clear that the probability of the coin being in a particular state at arbitrary timestep $k$ depends entirely on its most recent state, \emph{i.e.}, the process is completely characterized by the following conditional distributions:
\begin{align}\label{eq:pertcoin}
    &\mathbbm{P}_{k}(\texttt{H}_{k}|\texttt{H}_{k-1}) = \mathbbm{P}_{k}(\texttt{T}_{k}|\texttt{T}_{k-1}) = p \\
    &\mathbbm{P}_{k}(\texttt{H}_{k}|\texttt{T}_{k-1}) = \mathbbm{P}_{k}(\texttt{T}_{k}|\texttt{H}_{k-1}) = 1-p. \notag
\end{align}
The dependence of the future statistics on only the most recent outcome dramatically simplifies the complexity of any algorithm aiming to predict the behavior of this process. Instead of estimating the exponentially many probability distributions corresponding to sequences of outcomes over the entire course of history, one can simply condition on the previous state~\cite{Crutchfield1989,Shalizi2001,Crutchfield2011,Gu2012}. 

The type of process described above is known as a \emph{Markov} or \emph{Markovian process}, and is sometimes referred to as \emph{memoryless}, since the process itself stores no memory of historic states; the only temporal correlations that can arise are mediated through the state of the system itself. One can generalize the perturbed coin to incorporate longer memory effects by shaking the card harder each time the same outcome is observed and resetting the shaking strength as soon as an outcome differs from the previous one. In this case, although the statistics of the next state only depend upon the most recent sequence of outcomes, this does not imply an absolute demarcation of the process into some timesteps of memory and an irrelevant history. Indeed, temporal correlations between observations can be exhibited over various timescales; if one begins such a process with the coin facing \texttt{H} up, a few steps later it is more likely than not to be found in the same state. The crucial point is that, \emph{once we know} the state at timestep $k$, we may as well discard any observations of previous states, since they tell us no additional information. 

One accounts for genuine memory effects of this type that are finite in duration through the notion of \emph{Markov order}, $\ell$, which dictates that the statistics observed at any given time only depend upon knowledge of the past $\ell$ outcomes. In other words, with respect to knowledge of the state over a sequence of $\ell$ timesteps $\{k-\ell, \hdots, k-1\}$, any statistics of the states an experimenter might deduce over the history $\{ 1, \hdots, k-\ell-1 \}$ and the future $\{ k, \hdots, n\}$ timesteps are \textit{conditionally} independent. It is the act of \emph{observing a sequence of outcomes} that renders the future and history conditionally independent; one can think of this action as an intervention on the system that blocks any possible historic influence on the future dynamics. 

Formally, a (discrete-time, $n$-step) classical stochastic process is described by the joint probability distribution of the state of the system (represented as a random variable $X$ which takes values $x$) over the entire sequence of timesteps: $\mathbbm{P}_{n:1}(x_n, \hdots, x_1)$. A process has Markov order-$\ell$ when the distribution factorizes as
\begin{align}\label{eq:cmarkovorder}
    &\mathbbm{P}_{n:1}(x_n, \hdots, x_1) \\
    &= \prod_{j=\ell+1}^{n} \mathbbm{P}_{j}(x_j | x_{j-1} \hdots, x_{j-\ell}) \mathbbm{P}_{\ell:1}(x_\ell, \hdots, x_1), \notag
\end{align}
with the special case $\ell=1$ reducing to the condition of Markovianity. Again, $\ell$ determines the number of timesteps over which one must observe states in order to optimally predict, in principle, the next state, thereby providing a natural and fundamental timescale for memory length in stochastic processes. The practical importance of this property cannot be overstated, as processes with finite Markov order can be effectively reduced to Markovian processes upon a suitable grouping of timesteps, allowing for efficient simulation~\cite{StochProc,VanKampen1998}.

Implicit in this classical description is the assumption of the ability to observe realizations of the state at any time without affecting it. This immediately becomes problematic when attempting to characterize quantum stochastic processes, where this assumption simply cannot be satisfied: in quantum theory, measurements, in general, necessarily disturb the state. We now focus on a simple example of system-environment dynamics within the framework of generalized collision models with memory to give an intuitive understanding of the emergence of finite Markov order in quantum processes, before addressing the general setting.

\subsection{Generalized Collision Model with Memory}


\begin{figure*}
\centering
\includegraphics[width=0.9\linewidth]{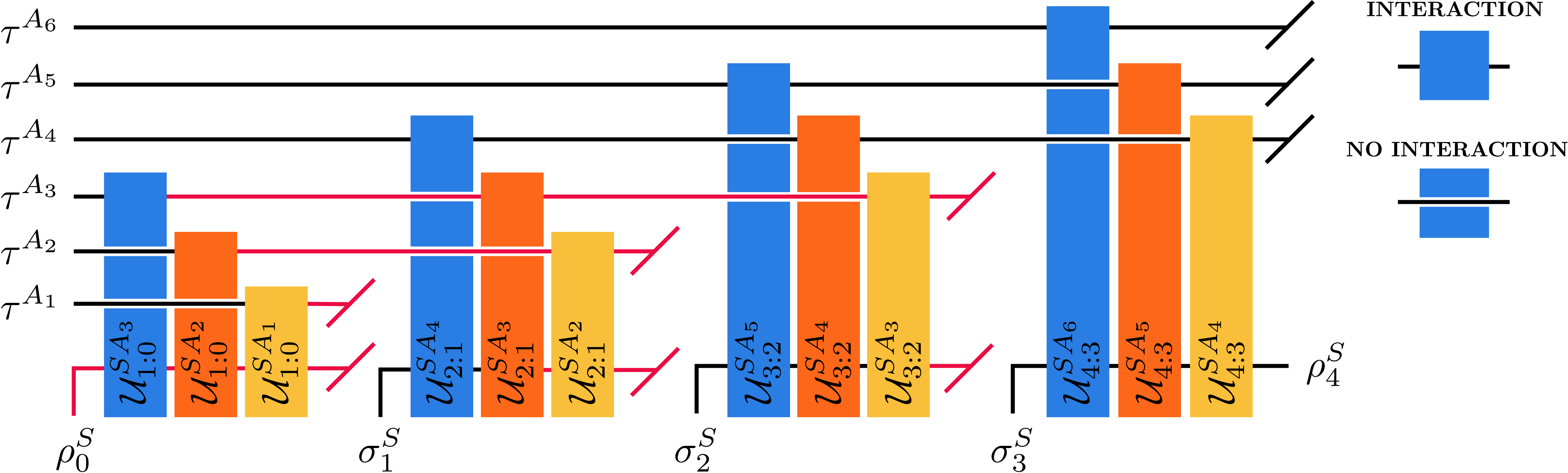}
\caption{\emph{Finite-length memory with respect to trash-and-prepare protocol.} The underlying system-environment dynamics for the generalized collision model described in the main text, interspersed with the trash-and-prepare protocol applied to the system. Any possible influence stemming from the history persists to impact the future for at most $\ell=3$ timesteps before being trashed. For instance, the red paths signify the degrees of freedom that can be affected by the initial preparation, whereas the black ones cannot be. The final state is a function of only the most recent $\ell$ preparations, $\{ \sigma_1^S, \sigma_2^S, \sigma_3^S \}$, and entirely independent of the initial system state, $\rho_0^S$. Any other instrument sequence on the system, \emph{e.g.}, a measurement rather than a trash-and-prepare instrument, would open up a pathway for the initial state $\rho_0^S$ to influence the future state $\rho_4^S$. 
\label{fig:trashingdilation}}
\end{figure*}


Within the field of open quantum dynamics, collision models have been introduced to provide a concrete underlying mechanism describing the evolution of memoryless processes~\cite{Rau1963,Ziman2002,Brun2002,Scarani2002,Ziman2005}. In such models, a system interacts with an environment comprising independent ancillary subsystems through successive unitary collisions with each ancilla. Because each ancilla is only interacted with once, there is no way for the environment to act as a mechanism for memory transport by influencing future dynamics. 

One can generalize this setting to allow for non-trivial memory effects: the most common approaches include beginning with an initially correlated environment~\cite{Rybar2012, Bernardes2014}, allowing for ancilla-ancilla interactions~\cite{Ciccarello2013, McCloskey2014, Lorenzo2016, Cakmak2017, Campbell2018}, permitting repeated system-ancilla collisions~\cite{Grimsmo2015, Whalen2017}, or some type of hybrid approach~\cite{Kretschmer2016,Lorenzo2017A, Lorenzo2017,Strathearn2018}. Each one of these scenarios can be motivated through realistic physical origins that demand some reasonable assumptions~\cite{Ciccarello2017}; in all of them, the environment acts as a memory by storing information about previous system states to govern future evolution (see Fig.~\ref{fig:collisionmodel} for illustration). Here, we focus on a special case of such dynamics with repeated system-ancilla interactions, which has application in studying phenomena with substantial time-delays between repeated interactions, \emph{e.g.}, developing feedback-assisted process control protocols~\cite{Grimsmo2015, Whalen2017}.

Consider specifically the following $n$-step process, depicted as a quantum circuit in Fig.~\ref{fig:trashingdilation}. A system $S$ interacts with some inaccessible environment $E$, which comprises $n+\ell-1$ initially uncorrelated ancillary systems $\tau_0^{E} := \bigotimes_{x=1}^{n+\ell-1} \tau^{A_x}$. The overall joint system-environment dynamics between timesteps $j-1$ and $j$ is represented by the map defined as: $\rho_{j}^{SE} = \widetilde{\mathcal{U}}_{j:j-1} \rho_{j-1}^{SE} := \tilde{u}_{j:j-1} \rho_{j-1}^{SE} \tilde{u}_{j:j-1}^\dagger$. In this particular example, the joint evolution is broken up into an ordered sequence of pairwise collisions between the system and ancillary states of the environment as follows:
\begin{align}\label{eq:cmdynamics}
    \widetilde{\mathcal{U}}_{j:j-1} := \mathcal{U}_{j:j-1}^{SA_{j}} \hdots \mathcal{U}_{j:j-1}^{SA_{j+\ell-1}},
\end{align} 
where the superscripts label the systems involved in the interaction. Following the dynamics between timesteps $j-1$ and $j$, the specific ancilla $A_{j}$ will have interacted with the system $\ell$ times; it is then discarded and never involved in the future evolution. 

In this model, we have not allowed for any initial system-environment correlations or ancilla-ancilla collisions; this type of evolution describes a time-translationally invariant microscopic model for processes with memory, which propagates through the $\ell$ ancillas that feed-forward to act like a linear memory tape. By design, we can see how memory effects arise: ancilla $A_{x}$ can store information about the system, acquired during its first interaction through $\mathcal{U}^{SA_x}_{x-\ell+1:x-\ell}$, and use it to influence the future dynamics up until its final interaction with the system via $\mathcal{U}^{SA_x}_{x:x-1}$.

Suppose then that we wish to characterize such a process. To do so, in practice, we must measure realizations of the state of the system at each timestep. We immediately face the problem that any such measurement both conditions the state of the environment and directly affects the state of the system. This leads to different future dynamics dependent on both the measurement outcomes observed and the way in which they were measured. In this sense, an operational framework for characterizing quantum stochastic processes must allow for probing interventions on the level of the system. In contrast to the underlying process at hand, an experimenter is assumed to have complete control over these instruments, which we formalize in Section~\ref{sec:framework}. The appropriate question relevant to quantum processes is how one can actively block the effect of history on the future dynamics over a finite number of timesteps.

\subsection{Memory Length}
\label{subsec:memlen}

The representation of the process in Fig.~\ref{fig:trashingdilation} is particularly illuminating: we can see the possible ways in which information about the initial system state can perpetuate forward in time along connected paths originating from some point in the history (traced in red). For the particular collision model described above, a history-blocking strategy involves discarding the system state we receive from the process and re-preparing one of a known set of states to feed into the process over a sequence of $\ell$ timesteps. It is clear that upon applying such a sequence of \emph{trash-and-prepare} instruments, any possible path connecting the history to the future across $\ell$ timesteps is broken, thereby guaranteeing that the future evolution of the system is independent of anything that happened to it prior to the trash-and-prepare sequence. We say that the process has Markov order-$\ell$ with respect to this instrument sequence. In line with standard intuition, $\ell$ quantifies the number of timesteps over which an experimenter must act on the system in order to block any influence of its history on its future evolution. 

For concreteness, consider the first four timesteps of the generalized collision model dynamics described above, with $\ell=3$. The final state of the system after some initial state $\rho^S_0$ (which can depend, in general, on its entire history) evolves both uncontrollably due to the process and also in a controllable manner due to active application of the trash-and-prepare instrument sequence, and is given by:
\begin{widetext}
\begin{align}\label{eq:trashingoutput}
    \rho_{4}^S = \ptr{E}{\widetilde{\mathcal{U}}_{4:3} \ \sigma_{3}^S \ \ptr{S}{\widetilde{\mathcal{U}}_{3:2} \ \sigma_{2}^S \ \ptr{S}{\widetilde{\mathcal{U}}_{2:1} \ \sigma_{1}^S \
    \ptr{S}{ \widetilde{\mathcal{U}}_{1:0} \rho_0^S \otimes \tau^E_0  } }}},
\end{align}
\end{widetext}
where $\widetilde{\mathcal{U}}_{j:j-1}$ are defined as per Eq.~\eqref{eq:cmdynamics}, $\tau_0^{E} = \bigotimes_{x=1}^{n+\ell-1} \tau^{A_x}$, and the map $\sigma_{j}^S \text{tr}_{S}(\cdot)$ acts to discard the system at timestep $j$ and re-prepare it in some known state of our choosing, $\sigma_j^S$. 

In Appendix~\ref{app:cm}, we prove that this trash-and-prepare protocol indeed blocks any possible influence that the history can have on the future evolution. Specifically, we show that for the particular process introduced above, at arbitrary time $k$, all future states of the system after application of any length-$\ell$ sequence of trash-and-prepare instruments can be uniquely described as a function of only the $\ell$ recently prepared states, for any prior history [referring to Eq.~\eqref{eq:trashingoutput}, $\rho^S_{k} = f(\sigma^S_{k-1}, \hdots, \sigma^S_{k-3}) \, \forall \, \{ \sigma^S_{k-1}, \hdots, \sigma^S_{k-3}\}$ for $k = \{ 4, \hdots n \}$, with no dependence on any previous state of the system such as $\rho^S_0$]. This result implies that any possible statistics an experimenter might observe in the history and future are conditionally independent given any length-$\ell$ trash-and-prepare sequence beginning at arbitrary timestep $k-\ell$. Explicitly defining the trash-and-prepare sequence in terms of operations on the system as
\begin{align}\label{eq:trashprotocol}
    \mathcal{D}_k^\ell(\rho_{k-1}^S, \hdots, \rho_{k-\ell}^S) := \sigma_{k-1}^S \ptr{S}{\rho_{k-1}^S} \hdots \sigma_{k-\ell}^S \ptr{S}{\rho^S_{k-\ell}},
\end{align}
in a slight abuse of notation, we can write
\begin{align}\label{eq:mutualinfostatement}
    I(\{n, \hdots, k \} : \{ k-\ell-1 , \hdots, 0\})_{\mathcal{D}^\ell_k} = 0.
\end{align}
By this, we mean that the mutual information between any possible statistics recorded on the future and history timesteps, which quantifies any possible correlation between them, vanishes for all length-$\ell$ trash-and-prepare sequences, $\mathcal{D}^\ell_k$. 

Conversely, having finite-length memory with respect to the trash-and-prepare protocol is a necessary but insufficient condition to deduce the system-environment model depicted in Fig.~\ref{fig:trashingdilation}. As an explicit counterexample, consider two timesteps of dynamics in which two ancillary states of the environment are initially entangled, represented by the density operator $\tau^{A_1 A_2}$, and in product with the initial system state $\rho_0^S$. The system interacts first with $A_1$ via $\mathcal{U}_{1:0}^{SA_1}$, before $A_1$ is discarded, and then with $A_2$ via $\mathcal{U}_{2:1}^{SA_2}$, with a trash-and-prepare instrument $\sigma^S_1 \text{tr}_S$ applied to the system in between. It is clear that the initial state $\rho_0^S$ can have no influence on the future evolution, since the final system state can be written uniquely as a map acting only on the preparation fed into the process, $\rho_2^S = \ptr{A_2}{\mathcal{U}_{2:1}^{SA_2} \sigma_1^S \otimes \tilde{\tau}^{A_2}}$, where $\tilde{\tau}^{A_2} := \ptr{S A_1}{\mathcal{U}_{1:0}^{SA_1} \rho_0^S \otimes \tau^{A_1 A_2}} = \ptr{A_1}{\tau^{A_1 A_2}}$ represents the reduced state of $A_2$ that, importantly, shows no memory of $\rho_0^S$. Therefore, the dynamics has finite Markov order $\ell=1$ with respect to the trash-and-prepare protocol, but evidently does not have the form depicted in Fig.~\ref{fig:trashingdilation}; namely, because the ancillas begin in an entangled, \emph{i.e.}, correlated, state. 

To summarize, in this section we have introduced a specific type of generalized collision model which, by construction, perpetuates information about the history via a particularly simple mechanism. This allows us to study explicitly how memory effects arise from the perspective of the underlying dynamics and build an intuitive understanding of the necessity for instrument-specific Markov order in quantum mechanics. The salient points to note are the following: \textbf{i)} The trash-and-prepare protocol does not block every type of memory. For arbitrary system-environment dynamics, following a length-$\ell$ trash-and-prepare sequence, $\rho_k^S$ (and all the future system states) will, in general, depend on both the known preparations $\{ \sigma^S_{k-1}, \hdots, \sigma^S_{k-\ell} \}$ and the previous historic states $\{ \rho^S_{k-\ell-1}, \hdots, \rho_0^S \}$. Thus, if one were to measure statistics on the future and history, one would see correlations, leading to a breakdown of Eq.~\eqref{eq:mutualinfostatement} and, hence, an appreciable memory effect. Throughout this Article, we provide various examples of processes that exhibit finite Markov order with respect to other sequences of instruments, but not this one. \textbf{ii)} Even for the special case of dynamics described above, application of a different sequence of instruments than the trash-and-prepare protocol would not lead to future dynamics that are independent of the history. For example, suppose that one were to perform a measurement at an intermediary timestep during a length-$\ell$ trash-and-prepare protocol. Here, the measurement would condition the state of the environment on its outcome, and hence the influence of the history can permeate through the memory block, leading to dependence of the final output on previous dynamics. Lastly, in Appendix~\ref{app:othercm}, we further explore some of the various other types of memory that can be naturally introduced into collision models.

From the considerations outlined above, it is clear that knowing the history-blocking sequence for a given process gives us information about the process at hand, but not necessarily all of it. Although we have made no assumptions on the action of the unitaries, the dynamics introduced above is a special case of generic quantum evolution and the trash-and-prepare protocol is just one of many possible sequences of instruments one might apply. For a taste of the possibilities, the dynamics can fit anywhere within the theory of open quantum systems and the history-blocking instruments can be generalized measurements, unitary operations, or even necessarily correlated in time. We now address the issue more broadly: given a process, what can we learn about the structure of its underlying dynamics through knowledge of a sequence of instruments that acts to erase the influence of the system's history on its future evolution? This question follows naturally from the general framework of Markov order for quantum processes, developed in Ref.~\cite{Taranto2018}.


\section{Framework}\label{sec:framework}


To formally introduce quantum Markov order, in this section we recap the process tensor formalism and show how it leads naturally to the instrument-specific notion of Markov order that is unavoidable in quantum mechanics (for a more thorough introduction, see, \textit{e.g.}, Refs.~\cite{Pollock2018A,Pollock2018L,Milz2017,Taranto2018}). Following this, we explore the structure of processes that satisfy the finite Markov order constraint, making no assumptions about the form of the underlying dynamics. We then highlight some of the non-classical memory effects that can arise in the quantum setting, such as those whose historic influence can be blocked only through application of unitary sequences or generalized quantum measurements, before narrowing in on processes with vanishing quantum conditional mutual information, of which classical stochastic processes with finite Markov order arise a special case. 

\subsection{Preliminaries}\label{sec:preliminaries}

\begin{figure}[t]
\centering
\includegraphics[width=\linewidth]{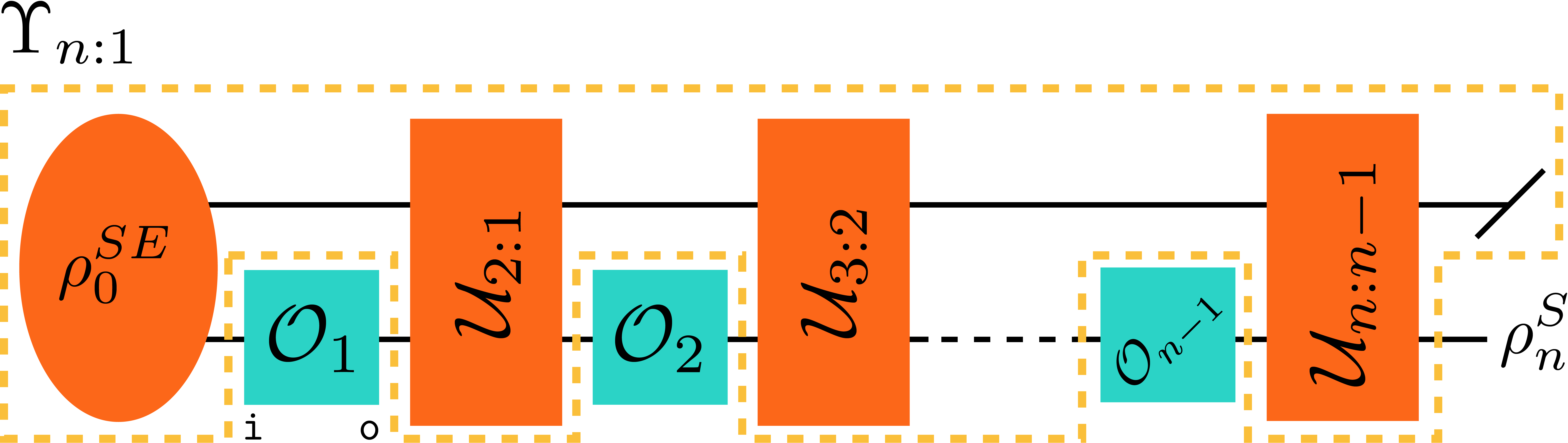}
\caption{\emph{Operational description of quantum stochastic processes.} Any quantum stochastic process can be modeled as arising from a system interacting unitarily with a suitably sized inaccessible environment. To characterize such evolution, one must interrogate the system. Hence, we allow for sequences of probing operations, $\mathcal{O}_j$ (green), to be applied to the system throughout the dynamics; these are the most general transformations allowable, taking input states to output states ($\inp, \out$ respectively). The final state of the system can be described uniquely through a multi-linear mapping on only the space of operations applied. This is precisely the process tensor, $\Upsilon_{n:1}$, represented as everything within the yellow, dashed boundary. \label{fig:stinespring}}
\end{figure}

We consider discrete-time processes on systems with a finite-dimensional state space. The state of the system $S$ at timestep $j \in \{ 1, \hdots , n\}$ is represented as an element of the bounded linear operators on a Hilbert space of dimension $d$: $\rho_j^S \in \mathcal{B}(\mathcal{H}_j)$. An operationally meaningful framework for stochastic dynamics necessarily consists of two parts, as shown in Fig.~\ref{fig:stinespring}: \textbf{i)} the \emph{uncontrollable} underlying process which governs the joint unitary evolution of the system with some inaccessible environment $E$, and \textbf{ii)} the interleaved \emph{controllable} changes to the state of the system, effected by our probing operations. In the generalized collision model of the previous section, the joint unitary evolution defined in Eq.~\eqref{eq:cmdynamics} provides the uncontrollable dynamics (the process), while the trash-and-prepare instruments defined in Eq.~\eqref{eq:trashprotocol} applied to the system induce changes to the state that we can control (our probing interventions). Such a description importantly provides a link between two distinct but equally valid perspectives of quantum stochastic dynamics: on the one hand, one can consider a process to be a black-box that can only be characterized with respect to statistics deduced by an experimenter probing the system; on the other hand, one can take the omniscient perspective of a being with knowledge of the deterministic underlying $SE$ dynamics. We now outline the most general setting possible that any stochastic evolution, quantum or classical, must fit within.

On the uncontrollable side, the environment need not begin uncorrelated from the system, nor be broken into ancillary states which interact with the system locally; all that is required is that the joint system-environment state evolves unitarily between timesteps. The evolution of the joint state from time $j-1$ to time $j$ is represented by the map $\mathcal{U}_{j:j-1} : \mathcal{B}(\mathcal{H}^{SE}_{j-1}) \to \mathcal{B}(\mathcal{H}^{SE}_{j})$ defined by $\rho^{SE}_{j} := \mathcal{U}_{j:j-1} \rho^{SE}_{j-1} = u_{j:j-1} \rho^{SE}_{j-1} u_{j:j-1}^\dagger$, where $u_{j:j-1}$ represents the unitary matrix corresponding to the joint evolution. Here, $\mathcal{U}_{j:j-1}$ can, in general and unlike in the previous example, act on the system and the \emph{whole} environment. 


On the controllable side, the trash-and-prepare protocol is simply a specific case of possible probing operations one might apply; more generally, these can be any physically realizable transformation in quantum mechanics, represented at each timestep $j$ by a completely-positive \textbf{(CP)} map $\mathcal{O}_j^{(x_j)} : \mathcal{B}(\mathcal{H}_j^\inp) \to \mathcal{B}(\mathcal{H}_j^\out)$, which take input system states to subnormalized output system states and whose trace is equal to the probability of realizing the outcome $x_j$, via: $\mathbbm{P}(x_j) \rho_j^\out = \mathcal{O}_j^{(x_j)} [\rho_j^\inp]$. The $\inp / \out$ labeling is, by convention, from the perspective of an experimenter implementing these operations to probe the process; we often refer to the input space as the space for \emph{measurements} and the output as the space for \emph{preparations}. The CP operation $\mathcal{O}_j^{(x_j)}$ describes how the state of the system is changed upon measuring outcome $x_j$, given that the \textit{instrument} $\mathcal{J}_j$ was used to interrogate the system. An instrument is any collection of such CP maps $\mathcal{J}_j = \{ \mathcal{O}_j^{(x_j)}\}$ that overall (\textit{i.e.}, when summed over) yield a completely-positive, trace-preserving \textbf{(CPTP)} map. More generally still, one could apply a sequence of instruments correlated across timesteps, \emph{e.g.}, by sending forward the ancilla that was used to implement an earlier operation. The corresponding transformations to the quantum system associated with observing a sequence of outcomes $x_{n-1:1}$ is captured by the multi-timestep CP map $\mathcal{O}_{n-1:1}^{(x_{n-1:1})}$. A collection of such maps which overall yields a valid quantum process (which we define shortly), is deemed a valid \emph{instrument sequence}. These represent the most general probing apparata one could implement over a sequence of timesteps (including no measurement at all) and have been formally introduced as \emph{testers} throughout the literature in the context of general quantum circuit architectures~\cite{Chiribella2009}.

In analogy to how one abstracts the environmental influence between two points in time as a quantum channel acting on the space of the system alone~\cite{Sudarshan1961,Jordan1961,Alicki1987,Davies1970}, one can abstract all that is uncontrollable in an open process across multiple timesteps as the \emph{process tensor}, $\Upsilon_{n:1}$, representing everything within the yellow, dashed line in Fig.~\ref{fig:stinespring}. This object is universal in the sense that it can describe any possible process permissible within quantum and classical physics~\cite{Chiribella2008-2,Chiribella2009,Pollock2018A}. The process tensor is a multi-linear functional that takes any sequence of CP maps as its input and outputs the resulting state of the system, subnormalized with respect to the probability of realizing the operation sequence in question, via the following rule:
\begin{align}\label{eq:processtensoraction}
    \rho_n^S &= \ptr{E}{\mathcal{U}_{n:n-1} \mathcal{O}_{n-1}  \hdots \mathcal{U}_{2:1} \mathcal{O}_{1} \rho_0^{SE}} \\
    &=: \ptr{n-1:1}{(\mathbbm{1}_n \otimes O_{n-1} \otimes \hdots \otimes O_1) \Upsilon_{n:1}}. \notag
\end{align}
To reiterate, here the controllable operations $\mathcal{O}_j$ act on the space of the system alone, while the joint unitary evolutions $\mathcal{U}_{j:j-1}$ act on the system and the inaccessible environment. It is evident that the process tensor contains information regarding the initial system-environment state and the successive joint unitary evolutions; we refer to this underlying system-environment model as a \emph{dilation} of the process. In contrast to the classical case, where a stochastic process is completely characterized by the underlying joint probability distribution over sequences of random variables representing observed outcomes, in quantum theory, each outcome corresponds to a CP map on the system. The process tensor provides the natural generalization of the joint probability distribution, encapsulating all possible multi-time correlations. 


\begin{figure*}[t]
\centering
\includegraphics[width=0.88\linewidth]{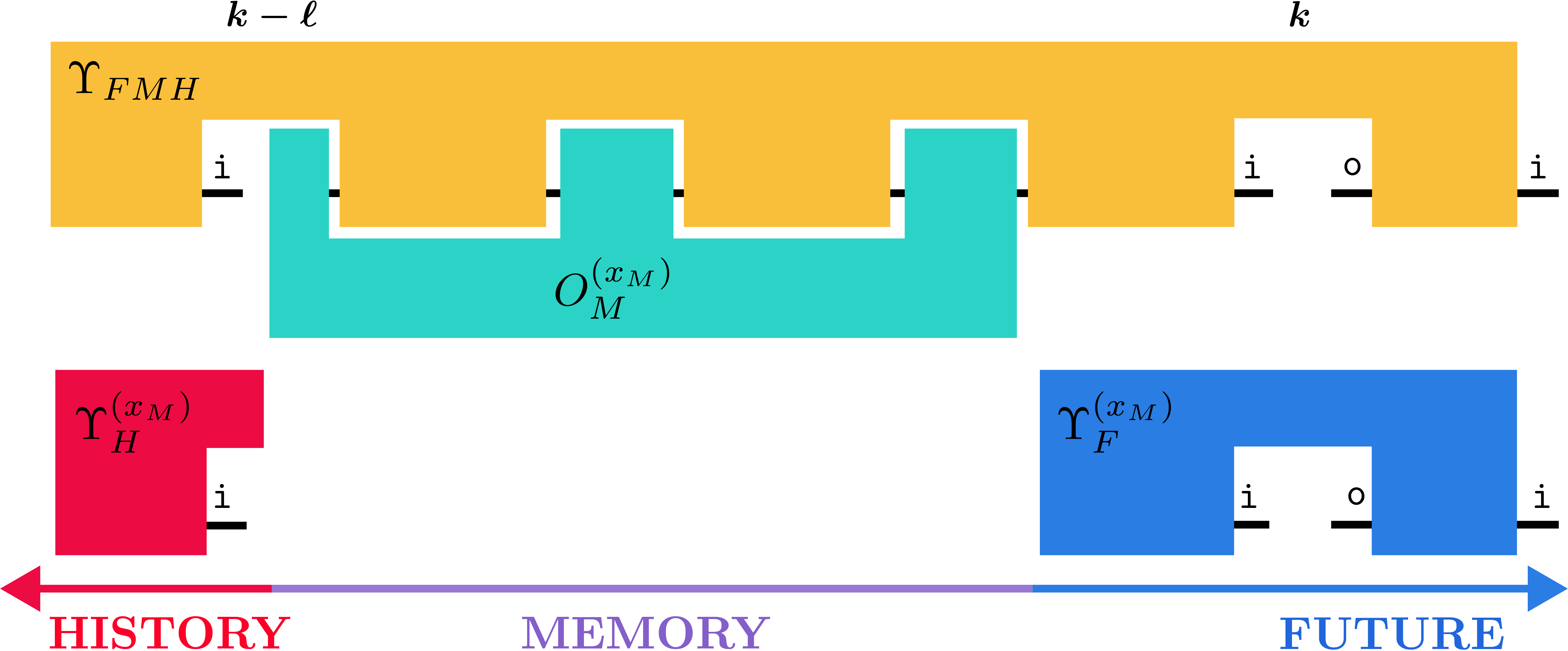}
\caption{\emph{Instrument-specific quantum Markov order.} An instrument sequence $\mathcal{J}_M$, comprising (temporally correlated) CP operations $\{ O_M^{(x_M)} \}$ (green) across a sequence of timesteps of length $\ell$, is applied to a process $\Upsilon_{FMH}$ (yellow). The process is said to have Markov order $\ell$ with respect to this instrument sequence if, for each possible realization of the instrument, $x_M$, the history (red, $\Upsilon_H^{(x_M)}$) and future (blue, $\Upsilon_F^{(x_M)}$) parts of the process are rendered conditionally independent. Here, for illustrative purposes, $M = \{ k-1^\out, \hdots, k-\ell^\out \}$; \emph{i.e.}, any possible measurement performed at timestep ${k-\ell}^\inp$ is considered part of the history, and its outcomes are conditionally independent of any future statistics. Eq.~\eqref{eq:genmutualinfostatement} calculates the mutual information between the conditional history and future processes, which vanishes for processes with finite Markov order. \label{fig:markovordercond}} 
\end{figure*}


In the second line of Eq.~\eqref{eq:processtensoraction} (and throughout this Article) we use an extended Choi-Jamio{\l}kowski isomorphism to represent the sequence of CP maps and the process tensor~\cite{Milz2017}, \textit{i.e.}, $O_{j}^\text{T} := (\mathcal{O}_j \otimes \mathcal{I} )[\Psi] \in \mathcal{B}(\mathcal{H}_j^\out\otimes \mathcal{H}_j^\inp)$ is the Choi state of the map $\mathcal{O}_{j}$, where $(\cdot)^\text{T}$ denotes transposition, $\mathcal{I}$ is the identity map and $\Psi := \sum_{xy} \ket{xx}\bra{yy}$ is an unnormalized maximally entangled state~\footnote{Here we define the Choi state as the transpose of its usual definition in order to ease notation, with no effect on any results.}. Similarly, $\Upsilon_{n:1} \in \mathcal{B}(\mathcal{H}_{n}^{\inp}\otimes \mathcal{H}_{n-1}^{\out}\otimes \hdots \otimes \mathcal{H}_1^{\inp})$ is a positive operator satisfying $\tr{\Upsilon_{n:1}} = \Pi_{j = 1}^n \text{dim}({\mathcal{H}_j^{\out}})$. Refer to Appendix~\ref{app:cji} for further details on the representation of the process tensor as a many-body Choi state and a summary of the labeling conventions used throughout this Article.

This isomorphism transforms temporal correlations into spatial ones, \emph{e.g.}, a Markovian process corresponds to an $\Upsilon_{n:1}$ of tensor product form~\cite{Pollock2018L,Pollock2018A}. Thus equipped, we can now apply standard correlation tools to understand properties of processes. It is important to note that \textit{all} processes can be represented in this way as (unnormalized) quantum states, but not all quantum states represent valid processes~\cite{Costa2018}. The set of possible temporal correlations are restricted, compared to their spatial counterparts, because the process tensor must satisfy a hierarchy of trace conditions which encode a proper causal-ordering, ensuring that the future process cannot influence the past~\cite{Chiribella2008, Chiribella2008-2, Chiribella2009,Milz2017, Milz2018, Pollock2018A}: 
\begin{align}\label{eq:causalconstraintmain}
    \ptr{j^\inp}{\Upsilon_{j:1}} = \mathbbm{1}_{j-1^\out} \otimes \Upsilon_{j-1:1} \quad \forall \; 1 < j \leq n.
\end{align}
Conversely, any positive, Hermitian operator satisfying this causality constraint is guaranteed to represent a valid \emph{temporal} evolution within quantum (and classical) theory~\cite{Chiribella2009,Pollock2018A}. Any valid instrument sequence must also satisfy a complementary set of trace conditions to be physically realizable. In summary, although all of the results presented throughout this Article are in terms of the Choi states of processes, these statements fundamentally address \emph{temporal} properties of processes, such as correlations between observables measured over time on some evolving quantum system.

Most importantly, the process tensor formalism allows one to calculate the joint statistics over the entire process, according to the following generalized spatio-temporal Born-rule~\cite{ShrapnelCosta2017},
\begin{align}\label{eq:processtensor}
    \mathbbm{P}_{n:1}(x_{n:1}|\mathcal{J}_{n:1}) = \tr{ O_{n:1}^{(x_{n:1})} \Upsilon_{n:1}},
\end{align} 
where we specify a measurement on the final output state. This allows us to unambiguously characterize important properties, such as Markovianity~\cite{Pollock2018A,Pollock2018L} and Markov order~\cite{Taranto2018}, of quantum stochastic processes from an operationally sound perspective. Since the process tensor framework contains the theory of classical stochastic processes as a special case, such characterizations reduce to the classical statement in the appropriate limit~\cite{Milz2017KET,Taranto2018}.

\subsection{Quantum Markov Order}

In this new language, the intuition behind the concept of Markov order remains unchanged from the standard one; we are still asking whether the future dynamics can be described completely, in principle, with knowledge accessible from the most recent $\ell$ states of the system. However, there are subtleties. In Ref.~\cite{Taranto2018}, we prove that demanding this constraint to hold for all possible instruments applicable to the system trivializes the theory into only admitting processes with Markov order $\ell=1$ or $\ell=\infty$. This leads naturally to the notion of instrument-specific Markov order, defined with respect to a particular choice of instrument sequence which acts to block the influence of the history on the future evolution. Equivalently, this instrument sequence renders the history and future parts of the process conditionally independent. In terms of the process tensor structure, the instrument-specific quantum Markov order condition implies that there exists an instrument sequence $\mathcal{J}_M = \{O_M^{(x_M)}\}$ such that the following holds at arbitrary timestep $k$ (see Fig.~\ref{fig:markovordercond} for illustration)~\cite{Taranto2018}:
\begin{align}\label{eq:qmarkovordercondition}
    \Upsilon_{FH}^{(x_M)} :=&
    \ptr{M}{ O_M^{(x_M)} \Upsilon_{FMH}}
    = \Upsilon_F^{(x_M)} \otimes \Upsilon_H^{(x_M)}
\end{align}
for all $O_M^{(x_M)}\in \mathcal{J}_M$, where, to ease notation, we group together timesteps as $\{ F, M, H\} := \{ \{ n , \hdots, k \}, \{ k-1, \hdots, k-\ell \}, \{ k-\ell-1, \hdots, 1 \} \}$. 

A few comments are in order. Firstly, if Eq.~\eqref{eq:qmarkovordercondition} is satisfied, we say that the process has Markov order $\ell$ with respect to the history-blocking instrument sequence, $\mathcal{J}_M$. A process can have finite Markov order with respect to entire families of instruments (as in the generalized collision model of Section~\ref{sec:collisionmodel}, where \emph{any} instrument of the form defined in Eq.~\eqref{eq:trashprotocol} is history-blocking). The fact that the process is rendered conditionally independent for each realization of the instrument, which is, overall, a \emph{deterministic} implementation, means that we are \emph{guaranteed} to block the effect of history upon application of the instrument in question (given that we know the outcome). In the more general setting, there may exist individual operation sequences that block the history; however, since these can only be implemented with some probability, in contrast to overall deterministic instrument sequences, such operations act to probabilistically render the future and history conditionally independent. In this Article, we focus on deterministic history-blocking sequences, where every constituent operation sequence in a collection that form a valid instrument sequence acts to block the effect of history. 

Secondly, satisfaction of Eq.~\eqref{eq:qmarkovordercondition} indeed guarantees the conditional independence of \emph{any} possible statistics one could obtain on the future and history given knowledge of the history-blocking instrument sequence. Generalizing Eq.~\eqref{eq:mutualinfostatement} to the case of an arbitrary history-blocking instrument sequence, we see that the mutual information between the conditional future and history processes for any realization of $\mathcal{J}_M$ vanishes, since they are of product form,
\begin{align}\label{eq:genmutualinfostatement}
    I(F : H)_{x_{M}} :=& S(\Upsilon_{F}^{(x_M)}) + S(\Upsilon_{H}^{(x_M)}) - S(\Upsilon_{FH}^{(x_M)}) \\ =& 0, \notag 
\end{align}
where $I(F : H)_{x_{M}}$ denotes the mutual information between the history and future processes given that the operation sequence corresponding to outcome $x_{M}$ was realized, and $S(\cdot)$ is the von Neumann entropy~\footnote{Since entropies are only well-defined for normalized objects, any entropic quantity is calculated using the normalized process tensor, \emph{i.e.}, $\Upsilon / \tr{\Upsilon}$.}. The mutual information upper-bounds all possible correlations between arbitrary observables on $F$ and $H$, and thus its vanishing implies the temporal regions of the future and history are totally uncorrelated with respect to knowledge of $x_M$~\cite{Kull2018}.

Thirdly, note that the conditional future process is a proper process tensor by construction, while the conditional history process represents an element of a tester, since the realization of the instrument sequence on the memory amounts to a post-selection~\cite{Chiribella2008,Milz2018,Taranto2018}. Intuitively, this means that when all possible outcomes are summed over, the conditional history is described by a proper process tensor, \emph{i.e.}, a positive semi-definite Choi state satisfying Eq.~\eqref{eq:causalconstraintmain}; however, the individual tester elements need not obey the latter condition. In the special cases where the they do, the probability of realising the associated sequence of outcomes of the history-blocking instrument can be extracted from the conditional history process, as we do at some points throughout this Article.

Lastly, in terms of notation, it is important to distinguish which input and output spaces constitute a memory block of length $\ell$. Any such block may begin and end on either the input or output Hilbert spaces of timesteps $k-\ell$ and $k-1$ respectively (see Fig.~\ref{fig:markovordercond}). To ease notation, we refrain from labeling each of these cases distinctly; instead, we provide visual representations of each example considered throughout this Article in order to make clear how the memory block is defined. 

As an example, to build a basic understanding of what the quantum Markov order condition in Eq.~\eqref{eq:qmarkovordercondition} entails, we refer to the main result in Ref.~\cite{Pollock2018L}: a quantum process is Markovian if and only if \textbf{(iff)} it displays Markov order $\ell=1$ (where \emph{only} the space $\mathcal{H}_{k-1}^\out$ is considered part of the memory block) with respect to an informationally-complete set of preparations, \textit{i.e.}, $\mathcal{J}_M = \{ \sigma_{k-1^\out}^{(x)}\}_{x=1}^{d^2} $ such that the set of states prepared spans the operator space $\mathcal{B}(\mathcal{H}_{k-1}^\out)$. Upon specifying such a set of states to feed into the process (that are necessarily independent of any prior history, as they can be chosen freely by the experimenter) and implementing what is referred to as a \emph{causal break} in Refs.~\cite{Pollock2018A,Pollock2018L}, \emph{any} future state of the system can be described in terms of CPTP maps acting on $\{ \sigma_{k-1^\out}^{(x)} \}$ alone. Demanding this condition to hold for each timestep in turn implies that the process tensor $\Upsilon_{n:1}$ can be decomposed as a sequence of CPTP maps $\Lambda_{k^\inp k-1^\out}: \mathcal{B}(\mathcal{H}_{k-1}^\out) \to \mathcal{B}(\mathcal{H}_{k}^\inp)$ acting on an initial system state $\rho_{1^\inp}$~\cite{Taranto2018,Pollock2018L}:
\begin{align}\label{eq:markov}
    \Upsilon^{\text{Markov}}_{n:1} = \Lambda_{n^\inp n-1^\out} \otimes \hdots \otimes \Lambda_{2^\inp 1^\out} \otimes \rho_{1^\inp}.
\end{align}
Here, each $\Lambda_{k^\inp k-1^\out}$ completely determines any possible statistics one might observe in the future through its action on $\sigma_{k-1^\out}^{(x)}$, which is uncorrelated from any possible past observations encoded in the previous $\Lambda_{{k-1}^\inp k-2^\out} \otimes \hdots \otimes \Lambda_{2^\inp 1^\out} \otimes \rho_{1^\inp}$. 

This product structure is equivalent to the necessary and sufficient characterization of Markovianity proposed in Ref.~\cite{Pollock2018L} and reduces to the classical Markov condition in the appropriate limit. Note that, if the set of preparations is not informationally-complete, one cannot uniquely deduce the tensor product structure of Eq.~\eqref{eq:markov}. Furthermore, there are processes with quantum Markov order $\ell=1$ that are non-Markovian, \textit{e.g.}, where a single trash-and-prepare instrument blocks the influence of history, as seen in (the counterexample given in) Section~\ref{subsec:memlen}: there, the future and history are conditionally independent with respect to an instrument defined on $\mathcal{H}_{k-1}^\out \otimes \mathcal{H}_{k-1}^\inp$, rather than only the most recent preparation space $\mathcal{H}_{k-1}^\out$ as required to be Markovian. Indeed, there are processes whose history is blocked with a trash-and-prepare instrument, but whose future dynamics can be conditioned by the measurement outcome corresponding to a realization of a measure-and-prepare instrument, and are therefore non-Markovian.

Extending this line of investigation, we are now interested in what the satisfaction of Eq.~\eqref{eq:qmarkovordercondition} for a particular instrument sequence implies for the underlying structure of the process tensor. The remainder of this Article presents our results regarding this question, with associated example processes to build intuition regarding memory length in quantum processes. The examples considered are constructed in such a way as to highlight some key peculiar features of quantum Markov order, and their essence applies to processes more broadly.

\section{Quantum Processes with Finite Markov Order}\label{sec:qmemory}

As illustrated in Section~\ref{sec:collisionmodel}, the fact that the memory of a process is blocked by an instrument sequence does not completely determine the underlying process. It does, however, impose certain structural constraints on the process. We begin by outlining the most general structure a process with finite Markov order, with respect to some instrument sequence, must have, before focusing on an important case of interest: where the history-blocking sequence is informationally-complete. We present a representative example of such processes in each case.

\subsection{Structure of Quantum Processes with Finite Markov Order}
\label{sec:genstructure}

Our structural analysis will be based on the fact that the process tensor is multi-linear in its arguments. Any $\ell$-step operation it acts on can be considered as an element of a vector space $W := \mathcal{B}(\mathcal{H}_{k-1}^\out \otimes \hdots \otimes \mathcal{H}_{k-\ell}^\inp)$ of dimension $\text{dim}(W)=d^{4 \ell}$. As already mentioned, the only constraint on a set of operations that constitute an instrument sequence is that they sum to a valid process, \textit{i.e.}, they are positive, Hermitian operators and their sum yields a positive, Hermitian operator with the same causal ordering as the process tensor that acts on them (this is enough to guarantee their physicality). This implies that the CP operations constituting an instrument sequence need not span the entire space $W$, even if they are linearly independent. We call an instrument sequence that spans $W$ \emph{informationally-complete} \textbf{(IC)}; such a sequence must contain a minimum number of $\text{dim}(W)$ linearly independent elements. On the other hand, we refer to an instrument sequence that does not entirely span $W$ as \emph{informationally-incomplete}.

Note that informational-completeness and history-blocking are two distinct properties of an instrument sequence. In particular, an informationally-incomplete instrument sequence can block the history, \emph{e.g.}, the trash-and-prepare sequence introduced in Section~\ref{subsec:memlen}. Informational-completeness pertains to whether or not one can completely characterize the process at hand through knowledge of its action on each constituent operation sequence. This property is of importance in this Section, which aims to identify structure in the process tensor given that we know a certain instrument sequence blocks the history.

We focus first on the most general case, where one has satisfaction of Eq.~\eqref{eq:qmarkovordercondition} for an arbitrary (potentially informationally-incomplete) history-blocking instrument sequence. It is sufficient to consider the subset of history-blocking instrument sequences comprising only linearly independent operations, since these provide the maximal amount of information one can obtain about the process. Suppose we have an informationally-incomplete history-blocking sequence $\mathcal{J}_M = \{ O_M^{(x)}\}_{x=1}^c$, where $c < \text{dim}(W)$. We can always complete this instrument sequence by appending an additional collection of operations, \textit{i.e.}, construct $\mathcal{J}_M^\prime = \mathcal{J}_M \cup \overline{\mathcal{J}}_M := \{ \{ O_M^{(x)} \}_{x=1}^c, \{ \overline{O}_M^{(y)}\}_{y=c+1}^{\text{dim}(W)}\}$. Since this entire collection of operations form a linearly independent set, there exists an associated \emph{dual set} of objects $\{ \Delta_M^{\prime (w)} \}$ such that $\tr{ O_M^{\prime (z)} \Delta_M^{\prime \dagger (w)}} = \delta_{zw} \, \forall \, z,w$~\cite{Modi2010, Modi2012A, Pollock2018A,Milz2017}~\footnote{Our definition of the dual set elements is related to that of Ref.~\cite{Modi2012A} by an adjoint, to make the connection with the Hilbert-Schmidt inner product explicit. Throughout this Article, we are always linearly expanding onto basis objects that are the Choi states to physical operations $\{ O^{(x)}\}$, which are Hermitian; in such cases, we have $\{ \Delta^{\dagger (y)} \} = \{ \Delta^{(y)} \}$. We therefore drop the adjoint symbol to ease notation.}. In terms of this (in general, non-orthonormal) basis, we can (completely) decompose the process tensor as: $\Upsilon_{FMH} = \sum_z^{\text{dim}(W)} \Upsilon_{FH}^{\prime (z)} \otimes \Delta_M^{\prime (z)}$. 

However, since we know that the instrument $\mathcal{J}_M$ acts to render the history and future into a tensor product for each outcome, we can further decompose the process tensor. We partition the total dual set into the elements dual to those in the history-blocking sequence, $\{\Delta_M^{(x)}\}_{x=1}^c$, and the rest, $\{ \overline{\Delta}_M^{(y)}\}_{y=c+1}^{\text{dim}(W)}$, such that $\tr{O_M^{(a)} \Delta_M^{(b)}} = \tr{\overline{O}_M^{(a)} \overline{\Delta}_M^{(b)}} = \delta_{ab}$ and $\tr{O_M^{(a)} \overline{\Delta}_M^{(b)}} = \tr{\overline{O}_M^{(a)} \Delta_M^{(b)}} = 0$. Now, the first $c$ terms in the sum above are given as $\sum_{x} \Upsilon_F^{(x)} \otimes \Delta_M^{(x)} \otimes \Upsilon_H^{(x)}$. Note that the duals in this construction are not necessarily positive operators, although the overall process must be. By direct insertion, it is clear that this portion of the process tensor indeed satisfies Eq.~\eqref{eq:qmarkovordercondition}. The remaining terms, which are inaccessible to the history-blocking operations, can be written as: $\sum_{y} \overline{\Upsilon}_{FH}^{(y)} \otimes \overline{\Delta}_M^{(x)}$. These encapsulate future-history correlations that an experimenter might observe upon application of an alternative instrument. This leads to the following theorem, which outlines the most general structure a process with finite quantum Markov order must have.

\begin{thm}\label{thm:markovorderstructure}
    Processes with finite quantum Markov order with respect to the instrument sequence $\mathcal{J}_M$ must be of the form
    \begin{align}\label{eq:markovorderstructure}
        \Upsilon_{FMH} =& \sum_{x=1}^{c}  \Upsilon_{F}^{(x)} \otimes \Delta_M^{(x)} \otimes \Upsilon_{ H}^{(x)} \\ &+ \sum_{y=c+1}^{\textup{dim}(W)}  \overline{\Upsilon}_{FH}^{(y)} \otimes \overline{\Delta}_M^{(y)}, \notag
    \end{align}
    where $c := |\mathcal{J}_M|$ is the number of constituent operations of the history-blocking instrument sequence, $\{ \Delta_M^{(x)} \} $ form the dual set to $\{ O_M^{(x)}\}$, satisfying $\textup{tr}{ \left[ O_M^{(x)} \Delta_M^{(y)} \right]} = \delta_{xy} \, \forall \, x, y$, and $\{\overline{\Delta}_M^{(y)}\}$ satisfy $\textup{tr}{ \left[ O_M^{(x)} \overline{\Delta}_M^{(y)} \right]} = 0 \, \forall \, x, y$. 
\end{thm}

Each term in the first summand has $F$ and $H$ in tensor product, ensuring Eq.~\eqref{eq:qmarkovordercondition} is satisfied with certainty for each realization of the instrument sequence in question. Such a decomposition must hold true for arbitrary timestep $k$ at which any length-$\ell$ memory block ends (although the terms in it can change for different blocks); a generic quantum process with infinite Markov order cannot be written so. The structure outlined highlights that for an informationally-incomplete history blocking sequence $\mathcal{J}_M$, we can only access a portion of the process tensor with conditionally-independent future and history. The $\overline{\Upsilon}_{FH}^{(y)}$ in the second term represents the portion of the process that can only potentially be revealed through other probing sequences. 

The generalized collision model explored in Section~\ref{sec:collisionmodel} is an example of such a process, since the trash-and-prepare protocol that blocks the effect of history constitutes an informationally-incomplete instrument sequence. This instrument sequence is, by its very nature, incoherent: an experimenter simply discards whatever states are output by the process and feeds in some of their own choosing, thereby blocking any possible effect of history on the future. In contrast to this, one might expect that applying sequences of coherent (\emph{i.e.}, unitary) operations to a process would always perpetuate memory effects from the history to the future by way of transmission through the system alone. We now provide an explicit counterexample: a process whose history is only blocked upon application of a sequence of coherent operations.


\begin{figure}[t]
\centering
\includegraphics[width=\linewidth]{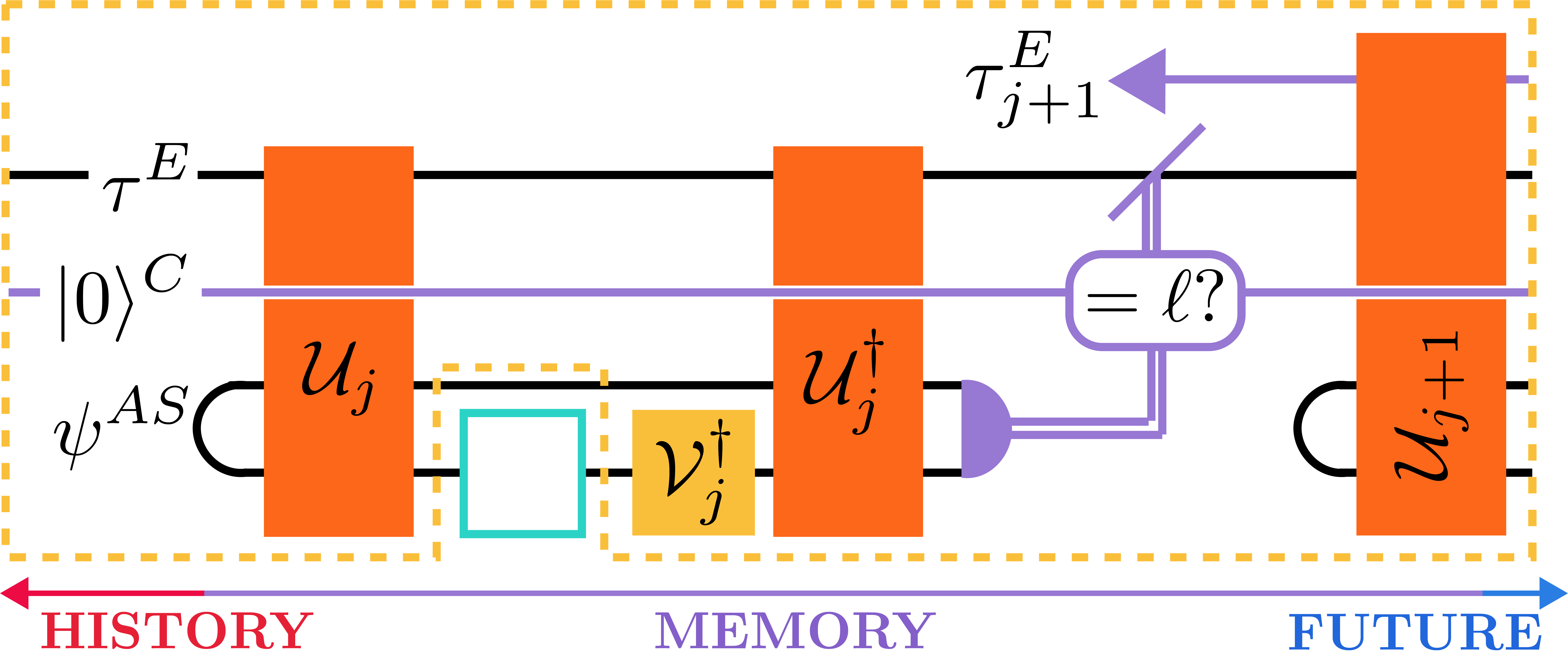}
\caption{\emph{Finite-memory with respect to a unitary instrument sequence.} The $SE$ dilation of a single timestep for a process whose historic influence on the future is blocked only by the sequence of unitary operations on the system $\{ \mathcal{V}_{k-\ell}, \hdots \mathcal{V}_{k-1}\}$. Everything inside the yellow, dashed boundary, including the unitary operation $\mathcal{V}_j^\dagger$, is part of the inaccessible process; we only have the choice of what operation is applied in the green outlined box. The ``cutting protocol'' described in the main text is depicted here in purple: the ancillary counter, $C$, registers the number of successive successful Bell basis measurements on the $SA$ system, which is re-prepared as a maximally entangled pair, $\psi^{AS}$, at each timestep. When the counter reaches $\ell$, the current environment state is discarded and a fresh one, $\tau_{j+1}^E$, is prepared to govern the future evolution. If the counter has not reached $\ell$, the environment is left to mediate correlations from the history to the future. \label{fig:unitary}}
\end{figure}


\begin{example} \label{ex:unitary}
\textit{History-Blocking with a Sequence of Unitaries.}---Consider the process depicted in Fig.~\ref{fig:unitary}. It is constructed such that there is exactly one length-$\ell$ sequence of unitary operations that guarantees the history is blocked, such that the Markov order of the process is equal to $\ell$. At each timestep $j \in \{ k-\ell, \hdots, k-1 \}$ in the memory block, the process prepares an ancillary system $A$ in a maximally entangled state with $S$, $\psi^{AS} := \tfrac{1}{d} \sum_{xy} \ket{xx}\bra{yy} $, which are together in tensor product with the rest of the environment $E$. 

The joint $EAS$ state undergoes dynamics according to some unitary map, $\mathcal{U}_j$, before an operation can be applied to the system $S$ by the experimenter. Following this operation, the process applies the inverse $\mathcal{V}_j^\dagger$ of some unitary map $\mathcal{V}_j$ on the system alone. Each of them is defined in terms of unitary matrices $v_j$ as $\mathcal{V}_j^\dagger(\rho_j^S) := v^\dagger_j \rho_j^S v_j$. The joint $EAS$ state then evolves according to the inverse unitary map $\mathcal{U}_j^\dagger$. Lastly, $AS$ is subject to the following ``cutting protocol'': a Bell basis measurement is implemented, with an ancillary system, $C$, counting whenever the outcome corresponds to $\psi^{AS}$. When $C$ reaches $\ell$, then the environment at that timestep is discarded, a fresh one is prepared to govern the future dynamics, and the counter is reset. If the correct measurement outcome is not observed, the environment is left untouched and the counter is also reset. 

It is evident that only upon application of the entire uncorrelated unitary sequence $\{ \mathcal{V}_{k-\ell} , \hdots, \mathcal{V}_{k-1}\}$ are the temporal correlations \emph{guaranteed} to be broken and the history and future processes rendered conditionally independent. If, on the other hand, this correct unitary sequence is not applied, the environment is allowed to mediate correlations between system states of the history and future, breaking the quantum Markov order condition. For any other sequence of operations implemented, while there is potentially a non-zero probability for the counter to reach $\ell$, this is not equal to 1; hence, overall, the influence of the history on the future is not blocked. This process is of the form of Eq.~\eqref{eq:markovorderstructure} with respect to the informationally-incomplete sequence of single-element (unitary) instruments, with the first sum containing a single term and the remainder of the process description encapsulated in the second term: 
\begin{align}\label{eq:unitarystructure}
    \Upsilon_{FMH} =& \frac{1}{d^\ell} \Upsilon^\prime_{F} \otimes V_{k-1}^\prime \otimes \hdots \otimes  V_{k-\ell}^\prime \otimes \Upsilon_{H} \\ &+ \sum_{y} \overline{\Upsilon}_{FH}^{(y)} \otimes \overline{\Delta}_M^{(y)}, \notag
\end{align}
where the $V_{j}^\prime / d$ are duals to the Choi states of the unitary maps $\mathcal{V}_j^\dagger$, and the conditional process tensor $\Upsilon^\prime_{F}$ is the fresh future process initiated by successful implementation of the cutting protocol. 
\end{example}

The process tensor in Eq.~\eqref{eq:unitarystructure} is evidently an example of Theorem~\ref{thm:markovorderstructure}; however, some remarks are in order. Firstly, note that even in the special case $\ell=1$, the process is non-Markovian, since it does not have the product structure outlined in Eq.~\eqref{eq:markov} (and the coherent unitary operation at timestep $k-1$ required to block the effect of history on the future operates on $\mathcal{H}_{k-1}^\out \otimes \mathcal{H}_{k-1}^\inp$). Secondly, no sequence of unitary operations can be IC; by definition, an informationally-incomplete sequence cannot be used to extract full information about a process. Although we know that any future dynamics will be independent of the history with respect to this sequence, we cannot predict what the next state will be as a function of the history-blocking sequence.

\subsection{History-Blocking with Informationally-Complete Instrument Sequences}\label{sec:infcomplete}

Interestingly, in Example~\ref{ex:unitary}, the influence of the history on the future is blocked only by a sequence of coherent (unitary) operations. This is somewhat counter-intuitive, as one might expect unitary transformations to perpetuate memory effects. In fact, the general structural constraint of Theorem~\ref{thm:markovorderstructure} is rather flexible, since knowledge of such an incomplete history-blocking instrument sequence does not determine the structure of the process at hand. In many cases of interest, one has access to an IC set of operations to probe the dynamics, \emph{e.g.}, when one attempts to tomographically reconstruct a generic process~\cite{Pollock2018A}. In this case, since an IC instrument sequence spans the entire space of operations, there can be nowhere for potential memory effects correlating the history and future to hide. The memory block can be decomposed onto an IC set of duals, uniquely specifying the entire process for each sequence of outcomes realized on the memory block. In this case, finding the future process to be conditionally independent of the history constrains the structure of the process tensor in a stricter manner than Eq.~\eqref{eq:markovorderstructure}; we immediately have the following corollary.

\begin{cor}\label{cor:complete} 
    A process with finite Markov order with respect to an informationally-complete history-blocking instrument sequence must have the following structure:
    \begin{align}\label{eq:markovorderstructure3}
        \Upsilon_{FMH} = &\sum_{x} \Upsilon_{F}^{(x)} \otimes \Delta_M^{(x)} \otimes \Upsilon_{ H}^{(x)}.
    \end{align}
\end{cor}

\begin{rem*}
    Theorem 4 in Ref.~\cite{Taranto2018} states that the only processes with finite quantum Markov order with respect to all instrument sequences are Markovian. Its proof begins by demanding Eq.~\eqref{eq:qmarkovordercondition} to hold for \emph{all} possible instruments. As such, we can consider an IC instrument sequence, in which case the process tensor must be of the form given by Eq.~\eqref{eq:markovorderstructure3}, which arises immediately from Theorem~\ref{thm:markovorderstructure}. Then, using the fact that one can construct arbitrary operation sequences spanning the operation space of $M$, we can vary $\Delta_M^{(x)}$ freely. Demanding the structure of Eq.~\eqref{eq:markovorderstructure3} to remain intact for arbitrary outcomes forces a tensor product between $M$ and $F$ or $H$ (or both), meaning the process tensor is restricted to a single term in Eq.~\eqref{eq:markovorderstructure3}, \textit{i.e.}, it is of product form. Requiring this to hold for any timestep leads to the Markovian product structure of Eq.~\eqref{eq:markov}. 
\end{rem*}

An operationally motivated choice for an IC instrument sequence consists of applying a \emph{causal break} at each timestep~\cite{Pollock2018L}: each operation here consists of an IC positive operator-valued measure \textbf{(POVM)} followed by an independent re-preparation of one of an IC set of states to feed forward at each timestep:
\begin{align}\label{eq:causalbreak}
    O_M^{(x_M)} = \bigotimes_{j=k-\ell}^{k} \sigma_{j^\out}^{(x_j^\out)} \otimes \Pi_{j^\inp}^{(x_j^\inp)}. 
\end{align}
Here, the $\{ \sigma_{j^\out}^{(x_j^\out)} \}$ are an IC set of states (\emph{i.e.}, they span the entire space of system states), the $\{ \Pi_{j^\inp}^{(x_j^\inp)} \}$ form an IC POVM satisfying $\sum_{x_j^\inp} \Pi_{j^\inp}^{(x_j^\inp)} = \mathbbm{1}_{j^\inp}$ and the $\{ x_j^\out, x_j^\inp \}$ are independent of each other. Such an operation acts to completely reset the state of the system and can be achieved, \emph{e.g.}, by making a measurement followed by a unitary operation, such that the output state is independent of the pre-measurement input. We now give an example of a process that exhibits finite Markov order with respect to an IC instrument sequence of causal breaks.

\begin{example}\label{ex:causalbreak}
\textit{History-Blocking by an Informationally-Complete Instrument Sequence (Causal Breaks)}---Consider the process depicted in Fig.~\ref{fig:infcomplete}, where, for simplicity, we present the case $\ell = 2$ for a three-step process, with the extension to longer length memory immediate. Initially, the following tripartite state is prepared:
\begin{align}\label{eq:icexamplestate}
    \rho_{Y 2^\inp 1^\inp} = \sum_{y} \mathbbm{P}(y) \rho_{Y}^{(y)} \otimes \Delta_{2^\inp}^{(y)} \otimes \rho_{1^\inp}^{(y)},
\end{align}
with $\{\Delta_{2^\inp}^{(y)}\}$ forming the dual set to some IC POVM $\mathcal{J}_{2^\inp} := \{\Pi_{2^\inp}^{(y)}\}$ and $Y$ labeling an ancillary Hilbert space of the environment that is never accessible to the experimenter. The marginal state $\rho_{1^\inp} := \ptr{Y 2^\inp}{\rho_{Y 2^\inp 1^\inp}}$ is fed out of the process at the first timestep, at which point the experimenter could implement any operation they choose; similarly, the state $\rho_{2^\inp}$ is fed out at the second timestep. The output states at $1^\out$ and $2^\out$ are mediated forward by the process, along with the $Y$ party of $\rho_{Y 2^\inp 1^\inp}$, as inputs to a CPTP map, whose Choi state is defined as follows:
\begin{align}\label{eq:icexamplemap}
    \Lambda_{3^\inp Y 2^\out 1^\out} := \sum_{xyz} \rho_{3^\inp}^{(xyz)} \otimes D_{Y}^{(y)} \otimes D_{2^\out}^{(z)} \otimes D_{1^\out}^{(x)},
\end{align}
where $\{D_{Y}^{(y)}\}$ are the dual set to $\{\rho_{Y}^{(y)}\}$, and $\{D_{2^\out}^{(z)}\}, \{ D_{1^\out}^{(x)}\}$ respectively form the dual set to some IC set of preparations $\{\sigma_{2^\out}^{(z)}\}, \{\sigma_{1^\out}^{(x)}\}$. This map acts to take each one of the $\sigma_{1^\out}^{(x)}, \rho_{Y}^{(y)}, \sigma_{2^\out}^{(z)}$ combination of inputs to one of $d^6$ unique states $\rho_{3^\inp}^{(xyz)}$, which are the final outputs of the process. 


\begin{figure}[t]
\centering
\includegraphics[width=\linewidth]{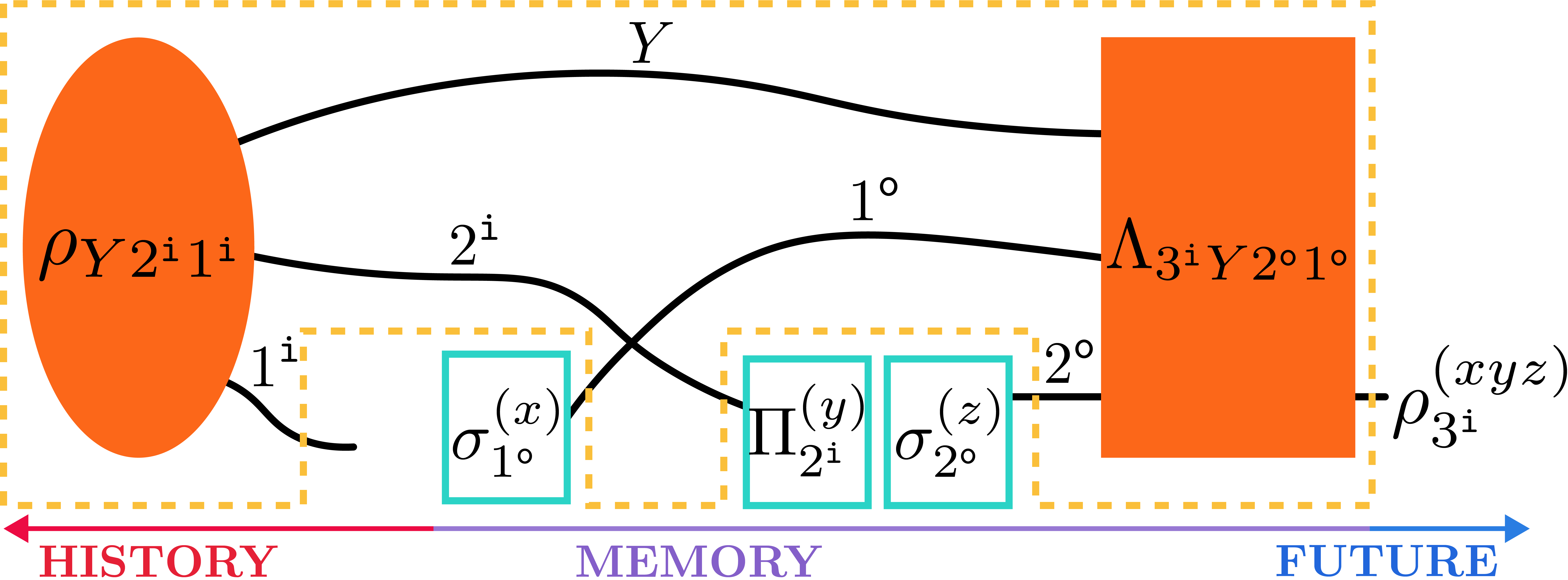}
\caption{\emph{Finite-memory with respect to an informationally-complete sequence.} Initially, a tripartite state $\rho_{Y2^\inp 1^\inp}$ is constructed as per Eq.~\eqref{eq:icexamplestate}, with subsystems $1^\inp, 2^\inp$ of it fed out at consecutive timesteps as described in the text. The states fed back into the process on spaces $1^\out, 2^\out$ are fed forward as inputs to the CPTP map $\Lambda_{3^\inp Y 2^\out 1^\out}$ defined in Eq.~\eqref{eq:icexamplemap}. Upon applying any combination of the correct IC causal break sequence $\{ \sigma_{1^\out}^{(x)}, \Pi_{2^\inp}^{(y)}, \sigma_{2^\out}^{(z)}\}$, one of $d^6$ final output states $\rho_{3^\inp}^{(xyz)}$ are output by the process in the future, each of which is conditionally independent of each historic $\rho_{1^\inp}^{(y)}$. If any other operations are applied, correlations will arise between the history and future in general.\label{fig:infcomplete}}
\end{figure}


Demanding the construction of $\rho_{Y 2^\inp 1^\inp}$ in Eq.~\eqref{eq:icexamplestate} to be a positive operator overall and the map $\Lambda_{3^\inp  Y 2^\out 1^\out}$ defined in Eq.~\eqref{eq:icexamplemap} to represent a valid evolution requires sufficient mixedness of each $\rho_{1^\inp}^{(y)}$ and $\rho_{3^\inp}^{(xyz)}$; additionally, ensuring that $\sum_{xyz} D_{Y}^{(y)} \otimes D_{2^\out}^{(z)} \otimes D_{1^\out}^{(x)} = \mathbbm{1}_{Y 2^\out 1^\out}$ guarantees $\Lambda_{3^\inp Y 2^\out 1^\out}$ satisfies the necessary trace conditions. Importantly, all of these conditions outlined above can be achieved simultaneously. It then follows that there exists an underlying unitary dilation of the map $\Lambda_{3^\inp  Y 2^\out 1^\out}$ that can be implemented in principle. The process tensor is explicitly given by
\begin{align}\label{eq:ptinfcomp}
    \Upsilon_{3^\inp: 1^\inp} = \sum_{xyz} \mathbbm{P}(y) \rho_{3^\inp}^{(xyz)} \otimes D_{2^\out}^{(z)} \otimes \Delta_{2^\inp}^{(y)} \otimes  D_{1^\out}^{(x)} \otimes \rho_{1^\inp}^{(y)}.
\end{align}

Intuitively, the IC instrument sequence $\mathcal{J}_M = \{ \sigma_{1^\out}^{(x)}, \Pi_{2^\inp}^{(y)}, \sigma_{2^\out}^{(z)}\}$ blocks any influence from the history to the future, as the measurement performed at $2^\inp$ leaves the initial state $\rho_{Y 2^\inp 1^\inp}$ in a product between $Y$ and $1^\inp$ for each outcome, such that the final output state is independent of any operation that could be performed at $1^\inp$. Indeed, for any realization of the instrument sequence, the conditional future and history processes are independent and of the form of Eq.~\eqref{eq:qmarkovordercondition}:
\begin{align}
    &\ptr{2^\out 2^\inp 1^\out}{\left( \sigma_{2^\out}^{(z)} \otimes \Pi_{2^\inp}^{(y)} \otimes \sigma_{1^\out}^{(x)}\right) \Upsilon_{3^\inp :1^\inp}}  \\
    &= \mathbbm{P}(y) \rho_{3^\inp}^{(xyz)} \otimes \rho_{1^\inp}^{(y)}. \notag
\end{align}

In this sense, the map $\Lambda_{3^\inp  Y 2^\out 1^\out}$ has no bearing on whether the effect of history is blocked or not: an experimenter could coarse-grain over any of the preparations while applying the correct measurement, \textit{e.g.}, feed in $p \sigma_{1^\inp}^{(x)} + (1-p) \sigma_{1^\inp}^{(x')}$, yielding a future state $p \rho_{3^\inp}^{(xyz)}+(1-p) \rho_{3^\inp}^{(x'yz)}$ that remains conditionally independent of the history $\rho_{1^\inp}^{(y)}$ given the measurement outcome $y$ at $2^\inp$. Of course, simpler processes can lead to an independent history and future with respect to the outcomes of an IC POVM (such as the example given in Appendix C of Ref.~\cite{Taranto2018}). However, here we construct a more general process with $\Lambda_{3^\inp  Y 2^\out 1^\out}$ defined as per Eq.~\eqref{eq:icexamplemap} in order to yield $d^6$ \emph{distinct} future states $\rho_{3^\inp}^{(xyz)}$ for each possible realization of the causal break sequence, each of which is conditionally independent of the history. We leave as an open problem the question whether there exist processes with finite Markov order with respect to an IC instrument sequence that are not causal breaks, such that each realization of the instrument leads to a distinct history and future process: \emph{i.e.}, Eq.~\eqref{eq:qmarkovordercondition} is satisfied for some IC $\mathcal{J}_M = \{ O_M^{(m)}\}$ such that $\Upsilon_F^{(m)} \neq \Upsilon_F^{(m')}$ and $\Upsilon_H^{(m)} \neq  \Upsilon_H^{(m')} \, \forall \, m, m'$.

Just as in the generalized collision model of Section~\ref{sec:collisionmodel}, in principle one can predict the next state of the system as a function of measurements and preparations in the causal break sequence. Furthermore, since the history-blocking sequence is IC, one can perform a process tomography to completely characterize the process as per Eq.~\eqref{eq:ptinfcomp}. If, on the other hand, one were to apply a different instrument on the memory block, then correlations between the future and history would in general arise (but, as already mentioned, we could vary the preparations and not see any influence from the history). 
\end{example}

\subsection{Summary of Section~\ref{sec:qmemory}}

The examples provided throughout Section~\ref{sec:qmemory} highlight significant properties of memory in quantum processes. Example~\ref{ex:unitary} explicitly shows that there exist processes where coherent (unitary) operations can break all possible temporal correlations between future and history, while Example~\ref{ex:causalbreak} highlights that the operations of a history-blocking instrument sequence can comprise an IC non-orthogonal set of independent measurements and preparations. 

So far, through Theorem~\ref{thm:markovorderstructure} and Corollary~\ref{cor:complete}, this section has developed the structural constraints that a process tensor must satisfy in order to exhibit finite quantum Markov order. However, this characterization is difficult to check in practice, due to the non-uniqueness of possible decompositions. It is therefore natural to seek a function of these finite Markov order processes that vanishes iff there are no correlations between the history and future remaining once a memory block of length $\ell$ is specified. For classical stochastic processes (without interventions), it is straightforward to show that the \emph{conditional mutual information} \textbf{(CMI)} of the underlying joint probability distribution has the desired property. In contrast, in both of the above examples (and also in the generalized collision model of Section~\ref{sec:collisionmodel}), the quantum generalization of the CMI evaluated on the Choi state of the process tensor between the history and future with respect to the memory is non-vanishing. This observation is insightful for a number of reasons which we address in the coming section, where we explore in detail the necessary conditions on the history-blocking instrument sequences for processes with vanishing quantum CMI, of which classical processes with finite Markov order are a special case. 


\section{Quantum Markov Order and the Quantum Conditional Mutual Information}\label{sec:vanishingqcmi}


We begin this section by briefly considering the relationship between Markov order and CMI in the classical setting. Any classical stochastic process with Markov order-$\ell$, \emph{i.e.}, described by a probability distribution satisfying Eq.~\eqref{eq:cmarkovorder}, can be equivalently characterized through the following two statements. Firstly, from an operational perspective, the significance of finite Markov order is best encapsulated through the existence of a so-called ``recovery'' map $\mathcal{W}_{M \to FM}$ that acts only on $M$ to give the correct future probability distribution: $\mathbbm{P}_{FMH} = \mathcal{W}_{M \to FM} [\mathbbm{P}_{MH}]$. This map can be straightforwardly used to simulate future dynamics, and the complexity of any predictive model is fundamentally bounded by the length of the block $M$ on which it acts (as well as by the number of possible realizations of each $X_j$). Secondly, an entropic characterization of finite Markov order that is easy to check in practice can be formulated as follows: the classical CMI vanishes $ I_\text{C}(F:H|M) := H(\mathbbm{P}_{FM}) + H(\mathbbm{P}_{MH}) - H(\mathbbm{P}_{FMH}) - H(\mathbbm{P}_{M}) = 0$, where $H(\mathbbm{P}_X) := - \sum_{x} \mathbbm{P}_X(x) \log{\mathbbm{P}_X(x)}$ is the Shannon entropy. Proving the equivalence between these statements is trivial: satisfaction of Eq.~\eqref{eq:cmarkovorder} implies the distribution factorizes as $\mathbbm{P}_{FMH}(x_F, x_M, x_H)  = \mathbbm{P}_{F}(x_F|x_M) \mathbbm{P}_{MH}(x_M, x_H)$; the recovery map $\mathcal{W}_{M \to FM}$ can be chosen to act as multiplication by the stochastic transition matrix $\mathbbm{P}_{F}(x_F|x_M)$. Equivalence to vanishing classical CMI is obvious by writing the conditional mutual information as a relative entropy between probability distributions (Kullback-Liebler divergence) in the following way, $I_\text{C}(F:H|M) = D^{KL}(\mathbbm{P}_{FH|M} \|\mathbbm{P}_{F|M} \mathbbm{P}_{H|M})$ and noting that this relative entropy vanishes iff the arguments are identical. Thus, in the classical setting, vanishing CMI is equivalent to finite Markov order.

As mentioned previously, until the recent introduction of the process tensor, there was no meaningful way to develop a sensible notion of Markov order, because the statistics that can be deduced depend on how one probes the process and are thus inherently instrument-dependent. Despite this concern, the alternative characterizations of Markov order in the classical setting described above can easily be generalized to the quantum realm. This has led to quantum Markov chains being defined throughout the literature as quantum \emph{states} with vanishing quantum CMI~\cite{Ruskai2002,Petz2003}, or, equivalently, those that satisfy quantum generalizations of recoverability~\cite{Hayden2004,Ibinson2008, Fawzi2015, Wilde2015, Sutter2016, Sutter2017}. On the other hand, the general theory of quantum Markov order for processes introduced here is captured by the conditional independence statement of Eq.~\eqref{eq:qmarkovordercondition}. This instrument-dependent statement is in stark contrast with the existing definitions on quantum states, which make no mention of the instrument sequence of choice. Therefore, it is unclear how such characterizations concretely relate to temporal processes with finite quantum Markov order. Nonetheless, the Choi-Jamio{\l}kowski isomorphism allows us to consider temporal processes in terms of their corresponding Choi state. We now explore the behavior of quantities used in analyzing spatial correlations in quantum states, such as the quantum CMI, for the Choi state of the process tensor and explore its relation to finite quantum Markov order. 

Understanding where processes with vanishing quantum CMI fit within our more general theory of finite quantum Markov order is also of significant practical interest. Unlike the classical case, proving the equivalence between quantum states with vanishing quantum CMI and those tripartite states that are recoverable through action on the conditioning subsystem alone is nontrivial and the proof is a highly celebrated result~\cite{Petz1986,Petz2003}. Another important result on the structure of such states arises by studying the set of states that remain unperturbed under action of such a recovery map~\cite{Koashi2002,Hayden2004}. In light of this, it is natural to seek processes whose future can, in principle, be simulated without disturbance through action of a map on the memory block alone. In addition, recent bounds on the fidelity of recovery for such states with approximately vanishing quantum CMI have been established~\cite{Fawzi2015,Sutter2016,Sutter2017,Wilde2015}, potentially providing a degree of confidence in approximately simulating processes with small memory effects. We have the following theorem that establishes a relation between vanishing quantum CMI and finite Markov order.


\begin{figure}[t]
\centering
\includegraphics[width=\linewidth]{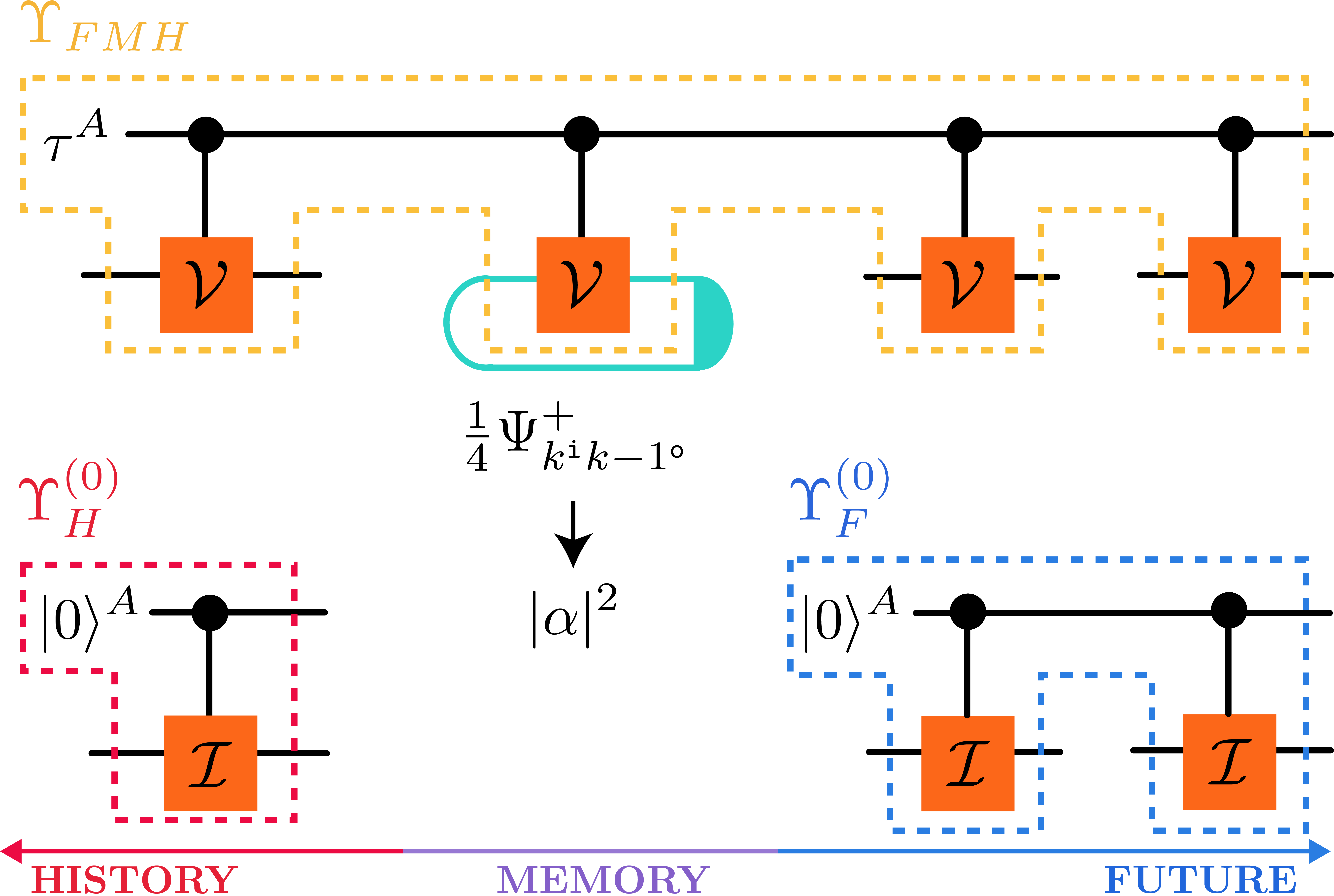}
\caption{\emph{Process with non-vanishing quantum conditional mutual information.} The environment is a four-dimensional ancilla. Its state is a coherent superposition of the basis states $\{\ket{0},\dots,\ket{3}\}$. The system-environment evolution is a control unitary, which implements one of the the four Pauli rotations $\mathcal{V} := \{ \mathcal{I}, \mathcal{X}, \mathcal{Y}, \mathcal{Z} \}$ on the system depending on the state of the ancilla (see top panel). The history-blocking instrument sequence consists of feeding in one half of a Bell pair and at the next measuring the system and the other half in the Bell basis at the next timestep. For each outcome of this instrument, one can infer which of the four Pauli rotations was applied, and the history and future processes are conditionally independent. For illustrative purposes, the bottom panel depicts the conditional processes that arise from successful implementation of the operation $\tfrac{1}{4} \Psi^+$, which occurs with probability $|\alpha|^2$. \label{fig:qcmi}}
\end{figure}


\begin{thm}\label{thm:cmi}
    Vanishing quantum CMI guarantees the process has finite quantum Markov order; the converse does not hold.
\end{thm}

We noted this statement as Proposition 5 in Ref.~\cite{Taranto2018}; here, we prove it concretely. The structure of processes with vanishing quantum CMI can be deduced from that of quantum states with vanishing quantum CMI, with the additional causality constraint imposed to ensure a valid process. The CMI of a quantum process is defined by $I(F:H|M) := S(\Upsilon_{FM}) + S(\Upsilon_{MH}) - S(\Upsilon_{FMH}) - S(\Upsilon_{M})$, and it vanishes iff there exists an orthogonal decomposition of the composite $M$ Hilbert space as $\mathcal{H}_M = \bigoplus_{m} \mathcal{H}_{M^L}^{(m)} \otimes \mathcal{H}_{M^R}^{(m)}$, such that~\cite{Hayden2004}
    \begin{align}\label{eq:qcmistates}
      \Upsilon_{FMH}^{CMI=0} &= \bigoplus_{m} \mathbbm{P}(m) \Upsilon_{F M^L}^{(m)} \otimes \Upsilon_{M^R H}^{(m)}.
    \end{align}
Here, the decomposition of $\mathcal{H}_M$ does not necessarily respect the temporal ordering of the underlying process. Specifically, the Hilbert spaces $\{ \mathcal{H}_{M^L}^{(m)} \}$ do not need to describe events that occur strictly before or after those described in $\{ \mathcal{H}_{M^R}^{(m)} \}$.

The proof of Theorem~\ref{thm:cmi} is given in Appendix~\ref{app:proofcmi}; the basic strategy is to explicitly construct a history-blocking instrument sequence for processes of the form in Eq.~\eqref{eq:qcmistates} and show that this structure is a special case of Eq.~\eqref{eq:markovorderstructure}, implying that there exist processes with finite Markov order but non-vanishing quantum CMI. The history-blocking sequence we construct is, in fact, made up of the set of orthogonal projectors (which form a self-dual set) onto each of the $m$ subspaces in the decomposition above. This begs the question: do processes with finite Markov order with respect to an instrument sequence comprising only orthogonal projectors necessarily have vanishing quantum CMI? Here, we provide an explicit example to show that this is not the case; indeed, the relationship between Markov order with respect to projective operations and vanishing quantum CMI is a subtle one.

\begin{example}\label{ex:qcmi}
\textit{Process with Non-Vanishing Quantum CMI but Finite Markov Order for a Sequence of Rank-1, Orthogonal Projectors.}---Consider the process depicted in Fig.~\ref{fig:qcmi}. Begin with a four-dimensional ancilla qudit in a coherent superposition $\ket{\tau}^A =  \alpha \ket{0} + \beta \ket{1} + \gamma \ket{2} + \delta \ket{3}$ with $|\alpha|^2+|\beta|^2+|\gamma|^2+|\delta|^2 = 1$. Controlled on the state of this qudit, the process implements one of the four Pauli maps (including the identity map), $\mathcal{V} := \{ \mathcal{I}, \mathcal{X}, \mathcal{Y}, \mathcal{Z} \}$, on a qubit system. The Choi states of these operations are the (unnormalized) four Bell pairs:
\begin{align}\label{eq:unnormalizedbell}
    \ket{\Psi^\pm} &:=  \ket{00}\pm\ket{11}, \\
    \ket{\Phi^\pm} &:=  \ket{01}\pm\ket{10}. \notag
\end{align}
Suppose that the process continues for $n$ timesteps and, at the $n^\text{th}$ step, the ancilla is fed out with the system in order to retain the quantum features of the process. The corresponding process tensor is $\Upsilon_{n:1} = \ket{\Upsilon} \bra{\Upsilon}$, where
\begin{widetext}
\begin{align}\label{eq:ptqcmi}
    \ket{\Upsilon} :=& \alpha \ket{0}^A_{n^\inp} \otimes \ket{\Psi^+_{n^\inp n-1^\out}  \hdots  \Psi^+_{2^\inp 1^\out}}  + \beta \ket{1}^A_{n^\inp} \otimes \ket{\Phi^+_{n^\inp n-1^\out} \hdots  \Phi^+_{2^\inp 1^\out}} \\
    &+ \gamma \ket{2}^A_{n^\inp} \otimes \ket{\Phi^-_{n^\inp n-1^\out}\hdots  \Phi^-_{2^\inp 1^\out}}  + \delta \ket{3}^A_{n^\inp} \otimes \ket{\Psi^-_{n^\inp n-1^\out}  \hdots \Psi^-_{2^\inp 1^\out}}. \notag
\end{align}
\end{widetext}
Note that this is not a Markovian process [it is not of the product form of Eq.~\eqref{eq:markov}], nor is it a classical probabilistic mixture of such processes; rather, the process tensor is a pure state representing a coherent superposition of implementing sequences of the four Pauli maps.

Consider the instrument sequence where, at some timestep $k-1$, an experimenter inputs half of one of the Bell pairs, feeds the other half forward to the next timestep $k$, and then makes a Bell basis measurement (see Fig.~\ref{fig:qcmi}). The corresponding instrument consists of the Choi states $\mathcal{J}_{{k}^\inp {k-1}^\out} = \{ O_{{k}^\inp {k-1}^\out}^{(x)} \} := \tfrac{1}{4} \{ \Psi^+_{{k}^\inp {k-1}^\out}, \Phi^+_{{k}^\inp {k-1}^\out}, \Phi^-_{{k}^\inp {k-1}^\out}, \Psi^-_{{k}^\inp {k-1}^\out} \}$. Since all cross terms in $\Upsilon_{n:1}$ are orthogonal to any of these, for each outcome observed upon their application, the experimenter observes one of the following four conditional processes:
\begin{align}\label{eq:ptqcmicond}
    \Upsilon^{(0)}_{FH} &= \Psi^+_F \otimes \Psi^+_H, \quad
    \Upsilon^{(1)}_{FH} = \Phi^+_F \otimes \Phi^+_H, \\
    \Upsilon^{(2)}_{FH} &= \Phi^-_F \otimes \Phi^-_H, \quad
    \Upsilon^{(3)}_{FH} = \Psi^-_F \otimes \Psi^-_H, \notag
\end{align}
where $\Psi^+_F := \ket{0}^A_{n^\inp} \otimes \Psi^+_{n^\inp n-1^\out} \otimes \hdots \otimes \Psi^+_{{k+1}^\inp {k}^\out}, \Psi^+_H := \Psi^+_{{k-1}^\inp k-2^\out} \otimes \hdots \otimes \Psi^+_{2^\inp 1^\out} $, and the superscript label corresponds to each possible realization (\emph{e.g.}, the label $(0)$ corresponds to feeding in half of the state $\Psi^+/2$ and successfully measuring it, which occurs with probability $\mathbbm{P}(0|\mathcal{J}_M) = |\alpha|^2$, and similarly for the other quantities defined). 

Intuitively, once an outcome of the instrument described is observed, one can deduce which of the four control operations were applied to the system and hence the state of the ancilla (which collapses onto one of its computational basis states and does not change further). This means that the history and future processes are known with certainty and are, therefore, conditionally independent. In contrast, suppose one were to perform an incoherent operation, such as feeding in the maximally mixed state before averaging over all measurement outcomes at the subsequent timestep. In this case, the conditional future-history process is now a probabilistic mixture of the four control operations being applied, \emph{i.e.}, $\Upsilon_{FH} = \sum_x \mathbbm{P}(x) \Upsilon^{(x)}_{FH}$, with $\{\Upsilon^{(x)}_{FH}\}$ defined in Eq.~\eqref{eq:ptqcmicond} and $\mathbbm{P}(x) = \{ |\alpha|^2, |\beta|^2,|\gamma|^2,|\delta|^2\}$. Such a mixture of Markovian processes is non-Markovian due to the correlations between the future and history: indeed, in this case one could condition the future dynamics by performing certain operations in the history. 

A simple calculation shows that the quantum CMI between the history and future given the memory for the process tensor in Eq.~\eqref{eq:ptqcmi} does not vanish; rather, it is equal to the Shannon entropy of the distribution $\mathbbm{P}(x) = \{ |\alpha|^2, |\beta|^2,|\gamma|^2,|\delta|^2\}$. Lastly, note that had we chosen to discard the ancilla, rather than feed it out at the final timestep, the corresponding process tensor is a probabilistic mixture of sequences of the four Pauli maps applied, \emph{i.e.}, the projector of Eq.~\eqref{eq:ptqcmi} without any cross terms. In this case, the process tensor is of the form in Eq.~\eqref{eq:qcmistates} and the quantum CMI vanishes.

\end{example}

In summary, here we have an example of a process which has finite Markov order with respect to an instrument sequence comprising only rank-1, orthogonal projectors, but nonetheless has non-vanishing quantum CMI. As detailed at the beginning of this section, such a situation cannot occur for classical stochastic processes; there, as long as an experimenter can observe realizations of the process at hand \emph{sharply}, \emph{i.e.}, resolve $d$ mutually exclusive outcomes for a $d$-dimensional system (which can be represented by $d$ rank-1, orthogonal projectors), then the classical CMI \emph{must} vanish for a process with finite Markov order. Thus, the example presented here represents a fundamentally quantum mechanical memory effect with no classical analog. Interestingly, in the study of classical stochastic processes where one allows for \emph{fuzzy} measurements, \emph{e.g.}, a measuring device which coarse-grains over some of the outcomes observed (which can be represented by higher-rank, orthogonal projectors), a similar discrepancy between the Markov order \emph{of the underlying process} and the vanishing classical CMI \emph{of the statistics observed} arises, as we now explore.

\section{Classical Stochastic Processes with Fuzzy Measurements}\label{sec:classical}

As noted by van Kampen, ``a physical process\dots may or may not be Markovian, depending on the variables used to describe it''~\cite{VanKampen1998}; and the same is true for the Markov order. The existence of perceived memory effects fundamentally depends on our experimental abilities, both in quantum mechanics (where it is generally acknowledged), as well as in classical physics, where it is often forgotten. Indeed, the standard framework for studying classical stochastic processes assumes the ability to measure observations of the random variables describing the system sharply; it breaks down when one allows for fuzzy measurements, or, more generally, experimental interventions. Such invasive operations are at the core of the theory of classical causal modeling~\cite{Pearl} (which contains classical stochastic processes as a special case~\cite{Milz2017KET}): here, one is allowed to implement any probing operations that map probability distributions in the state space to other valid distributions. As in quantum mechanics, allowing for interventions in classical physics makes the Markov order of a process inherently instrument dependent. Consequently, a comprehensive characterization of memory effects is important from an operational perspective, in particular for the case where one may not be able to resolve measurements at a sufficient level of granularity~\cite{Siefert2003, Bottcher2006, Lehle2011}. 

We now explore how the characterization of classical stochastic processes is inherently \emph{instrument-dependent} when one allows for the possibility of fuzzy measurements. Begin by noting that Eq.~\eqref{eq:cmarkovorder} can be reformulated in terms of the following statement on the conditional statistics:
\begin{align}\label{eq:cmarkovorder2}
    \mathbbm{P}_{k}(x_k | x_{k-1}, \hdots , x_1) = \mathbbm{P}_{k}(x_k | x_{k-1} \hdots, x_{k-\ell}),
\end{align}
which must be satisfied at any timestep $k \in \{ \ell + 1, \hdots, n \}$. Suppose that, instead of measuring the random variable $X$, we can only measure some $Y$, which coarse-grains over a subset of the $x$ values, denoted $\overline{x}$. The conditional statistics of the outcomes observed $y$ can be explicitly written in terms of the fine-grained variable $x$ as
\begin{align}\label{eq:cmarkovorder3}
    \mathbbm{P}_{k}(y_k | y_{k-1}, \hdots , y_1) =& \frac{\mathbbm{P}_{k:1}(y_k , \hdots , y_1) }{\mathbbm{P}_{k-1:1}(y_{k-1} , \hdots , y_1) } \\
    =& \frac{\sum_{\overline{x}} \mathbbm{P}_{k:1}(x_k , \hdots , x_1) }{\sum_{\overline{x}} \mathbbm{P}_{k-1:1}(x_{k-1} , \hdots , x_1) } \notag \\
    \neq&  \mathbbm{P}_{k}(y_k | y_{k-1} \hdots, y_{k-\ell}). \notag
\end{align}
Thus, even if Eq.~\eqref{eq:cmarkovorder2} is satisfied for the random variable $X$, it is not necessarily so for $Y$, nor does the classical CMI of the $Y$ variables vanish. The fact that coarse-graining can increase the memory length observed by an experimenter arises from the well-known property that the space of Markovian processes is not convex. Interestingly, we can also have the opposite scenario occur, \emph{i.e.}, have a process display finite Markov order with respect to a fuzzy measurement sequence, but if one had the ability to realize sharp observations of the process, they would attribute to it a longer memory length. Explicit examples for each of these scenarios are given in Appendix~\ref{app:classical}, with a quantum mechanical analog of the latter provided in Appendix~\ref{app:quantumfuzzy}.

The instrument-specific definition of Markov order described in this Article unambiguously characterizes memory length in any case of classical processes with fuzzy measurements and/or experimental interventions. In this light, even in classical physics, we should say that if a classical process is considered to have Markov order-$\ell$, it does so \emph{with respect to sharp observations of the process}. Even in the most general classical setting of causal modeling, however, the instrument dependence of Markov order is liftable, in the sense that it can be removed by changing perspective. By incorporating the experimenter \emph{and} their choice of intervention into the description of the process, the standard definitions of Markov order apply on a higher level. On the other hand, in the study of quantum stochastic processes, even sharp quantum measurements \emph{look} fuzzy when they act on a general state; thus, the fuzzy-measurement issue is fundamentally unavoidable and must be acknowledged accordingly through an instrument-specific notion of Markov order. 

\section{Conclusion}

In this Article, we have aimed to outline some of the key features of memory effects, in particular, memory length, in quantum stochastic processes. We began with a motivating example in Section~\ref{sec:collisionmodel} that highlights how a certain generalized collision model exhibits finite-length memory with respect to a natural trash-and-prepare history-blocking protocol; the deeper exploration of memory effects in similar models in Appendix~\ref{app:collision} further motivates the necessity of instrument-specific Markov order for quantum processes and a better understanding of the microscopic mechanism for memory propagation. We then tackled the general problem: given a sequence of operations that acts to erase the memory of a process, what can we say about its structure? After introducing the necessary formalism in Section~\ref{sec:framework}, in Section~\ref{sec:qmemory} we detailed the generic structural constraint on process tensors with finite quantum Markov order, exhibited, \emph{e.g.}, by a process whose history is blocked by a sequence of unitary operations (Example~\ref{ex:unitary}). We then considered processes with finite memory with respect to IC instrument sequences such as an IC POVM followed by an independent re-preparation of a state from an IC set (Example~\ref{ex:causalbreak}). In Section~\ref{sec:vanishingqcmi}, we showed that, unlike the classical case, in the quantum realm, processes with finite Markov order with respect to a sequence of instruments need not necessarily have vanishing quantum CMI (as exhibited by all examples throughout this Article including each type of collision model with memory). We provided an explicit example where the history-blocking instrument sequence comprises only sharp, orthogonal projectors, but nonetheless has non-vanishing quantum CMI (Example~\ref{ex:qcmi}). This is a fundamentally quantum mechanical phenomenon which has no classical analog. In contrast, when one can observe realizations of classical processes sharply, finite Markov order and the vanishing of the classical CMI are equivalent statements. Indeed, this section highlights that even in the classical case, we must be more careful in how we model stochastic processes when we cannot assume that we can measure realizations of the process sharply, as explored in Section~\ref{sec:classical}. In contrast to the classical case, however, in quantum mechanics, the fact that perceived memory effects are inherently instrument-dependent is fundamentally unavoidable and must be reconciled, as we have here and in Ref.~\cite{Taranto2018}.

Our present work raises some interesting avenues for future exploration. Firstly, none of the structural constraints imposed by the finite quantum Markov order condition rely on the underlying unitary dynamics, and even the generalized collision models we have studied make no assumptions on the action of such unitaries; our statements hold in general. On the other hand, realistic physical scenarios are often modeled by specific forms of interactions, \textit{e.g.}, nearest-neighbor interaction spin chains evolving in a time-translationally invariant manner. In such a scenario, while a generic sequence of instruments such as the trash-and-prepare protocol does not always act to block the historic influence, in practice it may be the case that such a sequence almost always approximately blocks the influence of history. A natural extension to this work would involve a deeper exploration of memory effects in specific physical models with the instrument-specific quantum Markov order formalism. Secondly, here we have not addressed the important issue of quantifying memory strength or classifying processes with approximately finite-length memory. It is clear that, unlike the classical case, the quantum CMI is a poor quantifier of memory strength, since it does not necessarily vanish for processes with finite quantum Markov order. In future work, we aim to address this issue by proposing instrument-specific measures of memory strength for quantum processes. Lastly, a better understanding of the type of memory and resources required to simulate finite-memory processes is needed. From the entanglement structure of the process tensor, one should be able to deduce whether the memory required to simulate a process is quantum or classical in nature and, from a practical perspective, if one is attempting to design quantum circuits with finite-length memory, the structure of the circuit must follow the constraints outlined in this Article.

We now move to discussing some of the broader implications of our work. It is clear that the process tensor provides the most generic description of causally-ordered processes allowable within quantum theory and thereby enables unambiguous characterization of complex time evolution. Examining properties of its structure, as we have in this Article, provides fundamental insight into understanding the space of quantum processes and temporal correlations. Indeed, similar objects that are not necessarily causally-ordered, such as the process matrix, are developing into a tool for studying the most general spatio-temporal correlations allowable~\cite{Oreshkov2012,Oreshkov2016,Costa2018}, shedding light on the defining features of quantum and post-quantum theories. On the practical side, the process tensor contains all the information one could ever hope to learn about a process. This, unfortunately, makes it computationally daunting to approach. In light of this, its usefulness lies in our ability to develop compression and extraction methods to approximate interesting physical evolutions with overlapping process tensors of finite length for efficient simulation of long-term dynamics. Indeed, this is the flavor of many methods proposed throughout the literature, such as the transfer tensor approach~\cite{Cerrillo2014,Rosenbach2016,Kananenka2016,Pollock2018T}. A deeper understanding of this will naturally begin to answer questions such as the following: \textit{How can processes be optimally compressed to reduce complexity in storing and simulating them? What resources are required for their simulation? How do errors accumulate if we try to keep reconstructing and disturbing overlapping parts of a process?} These, among others, have significant consequences for efficient quantum simulation and computation.


\begin{acknowledgments}

P. T. thanks F. Ciccarello, S. Lorenzo, and G. M. Palma for hosting a productive and enjoyable stay in Palermo which contributed in part to the present work, and M. Korenkova for thoughtful comments. All the authors thank M. Tomamichel and T. Notoh for insightful discussion, C. Jacobs for asking the right questions, and J. Morris for inspiring the protocol introduced in Section~\ref{sec:collisionmodel}. P. T. is supported by the Australian Government Research Training Program (RTP) Scholarship and the J. L. William Scholarship. S. M. is supported by the Monash Graduate Scholarship (MGS), Monash International Postgraduate Research Scholarship (MIPRS), and the J. L. William Scholarship. K. M. is supported through the Australian Research Council Future Fellowship FT160100073.

\end{acknowledgments}


\onecolumngrid
\appendix

\section{Generalized Collision Models with Memory}\label{app:collision}

\subsection{Generalized Collision Model with Memory via Repeated System-Ancilla Interactions}\label{app:cm}

In Section~\ref{sec:collisionmodel}, we introduced a type of underlying system-environment dynamics that arises from a generalized collision model, where the system interacts $\ell$ times with each ancilla in the order depicted in Fig.~\ref{fig:trashingdilation}. We claimed that the state of the system subject to such dynamics interspersed with the application of $\ell$ trash-and-prepare operations can be expressed as a function of only the last $\ell$ preparations. In this appendix, we explicitly prove this statement. 

Consider, without loss of generality, the case for $\ell=2$ (the extension to larger $\ell$ is straightforward). The final output state of the system following two trash-and-prepare instruments, with the re-preparations of the system state at time $j$ represented by $\sigma_j^S$, is given by 
\begin{align}
      \rho_3^S &= \ptr{A_4 A_3}{\mathcal{U}_{3:2}^{SA_3} \mathcal{U}_{3:2}^{SA_4} \sigma_2^S \ptr{S A_2}{\mathcal{U}_{2:1}^{SA_2} \mathcal{U}_{2:1}^{SA_3} \sigma_1^S \ptr{S A_1}{\mathcal{U}_{1:0}^{SA_1} \mathcal{U}_{1:0}^{SA_2} \rho_0^S \otimes \tau^{A_1} \otimes \tau^{A_2} \otimes \tau^{A_3} \otimes \tau^{A_4}}}} \\
      &= \ptr{A_4 A_3}{\mathcal{U}_{3:2}^{SA_3} \mathcal{U}_{3:2}^{SA_4} \sigma_2^S \otimes \tau^{A_4} \ptr{S A_2}{\mathcal{U}_{2:1}^{SA_2} \mathcal{U}_{2:1}^{SA_3} \sigma_1^S \otimes \tau^{A_3} 
      \ptr{S A_1}{\mathcal{U}_{1:0}^{SA_1} \tau^{A_1} \otimes \mathcal{U}_{1:0}^{SA_2} \rho_0^S \otimes \tau^{A_2}}}}. \notag
\end{align}
Now note that we can write the joint $SA_2$ state after the first interaction, \textit{i.e.},  $\mathcal{U}_{1:0}^{SA_2} \rho_0^S \otimes \tau^{A_2}$, as $\tilde{\rho}_0^S(\rho_0^S, \tau^{A_2}) \otimes \tilde{\tau}^{A_2}(\rho_0^S, \tau^{A_2})$, where $\tilde{\rho}_0^S(\rho_0^S, \tau^{A_2}) := \ptr{A_2}{\mathcal{U}_{1:0}^{SA_2} \rho_0^S \otimes \tau^{A_2}}$ and similarly for $\tilde{\tau}^{A_2}(\rho_0^S, \tau^{A_2})$. This simply expresses the post-interaction marginal states (marked with the tilde) as a linear map acting on the pre-interaction states. We do this in order to clearly track dependency of states through the process with respect to arbitrary unitary interactions. Continuing from above and repeatedly applying this method, we yield
    \begin{align}\label{eq:trashoutput}
      \rho_3^S &= \ptr{A_4 A_3}{\mathcal{U}_{3:2}^{SA_3} \mathcal{U}_{3:2}^{SA_4} \sigma_2^S \otimes \tau^{A_4} \ptr{S A_2}{\mathcal{U}_{2:1}^{SA_2} \tilde{\tau}^{A_2}(\rho_0^S, \tau^{A_2}) \otimes \mathcal{U}_{2:1}^{SA_3} \sigma_1^S \otimes \tau^{A_3}}} \\
      &= \ptr{A_4 A_3}{\mathcal{U}_{3:2}^{SA_3} \tilde{\tau}^{A_3}(\sigma_1^S, \tau^{A_3}) \otimes \mathcal{U}_{3:2}^{SA_4} \sigma_2^S \otimes \tau^{A_4}} \notag \\
      &= \ptr{A_4 A_3}{\mathcal{U}_{3:2}^{SA_3} \ptr{S}{\mathcal{U}_{2:1}^{SA_3} \sigma_1^S \otimes \tau^{A_3} } \otimes \mathcal{U}_{3:2}^{SA_4} \sigma_2^S \otimes \tau^{A_4} } \notag \\
      &= \mathcal{M}[\sigma_1^S, \sigma_2^S]. \notag
    \end{align}
Here, in the penultimate line, we re-expanded $\tilde{\tau}^{A_3}(\sigma_1^S, \tau^{A_3})$ to make explicit the fact that $\rho_3^S$ is a function of only the two previously prepared states, which can be written as a linear map $\mathcal{M}$ as in the final line, with no dependency on prior historic states such as $\rho_0^S$. Through time-translational invariance, the proof method holds for arbitrary timesteps and the extension to longer $\ell$ is immediate. Indeed, the process depicted in Fig.~\ref{fig:trashingdilation} has a length-$\ell$ memory with respect to the trash-and-prepare protocol. 

If, on the other hand, one were to apply a different instrument, then the output state here $\tilde{\rho}^S_3$ would in general show dependence on the historic state $\rho_0^S$. Consider for concreteness that an experimenter were to first apply a trash-and-prepare instrument and then at the second timestep a measurement on the system of some outcome $m$ followed by an independent re-preparation of the system into the state $\sigma_2^S$. Changing the second operation to a measurement and re-preparation amounts to introducing the local system measurement operator, $ \Pi_2^{(m)}$, in Eq.~\eqref{eq:trashoutput} following the joint unitary dynamics $\widetilde{\mathcal{U}}_{2:1}$, which leads to the following state at the third timestep:
    \begin{align}
      \rho_3^{\prime S} &= \ptr{A_4 A_3}{\mathcal{U}_{3:2}^{SA_3} \mathcal{U}_{3:2}^{SA_4} \sigma_2^S \otimes \tau^{A_4}  \ptr{S A_2}{ \Pi_2^{(m)} \left( \mathcal{U}_{2:1}^{SA_2} \tilde{\tau}^{A_2}(\rho_0^S, \tau^{A_2}) \otimes \mathcal{U}_{2:1}^{SA_3} \sigma_1^S \otimes \tau^{A_3} \right) \Pi_2^{(m)} }}.
    \end{align}
However, since the system and ancillas $A_2$ and $A_3$, in general, build up correlations during the interactions $\mathcal{U}_{2:1}^{SA_2}$ and $\mathcal{U}_{2:1}^{SA_3}$, the ancillary state $\tilde{\tau}^{A_3}(m)$ that feeds forward into the next step of dynamics will be conditioned upon the measurement outcome $m$, which implicitly depends upon the initial system state $\rho^S_0$; indeed, the future dynamics proceeds differently for distinct histories. Explicitly, we can write:
    \begin{align}
      \rho_3^{\prime S} &= \ptr{A_4 A_3}{\mathcal{U}_{3:2}^{SA_3} \tilde{\tau}^{A_3}(m; \rho_0^S, \sigma_1^S, \tau^{A_2}, \tau^{A_3}) \otimes \mathcal{U}_{3:2}^{SA_4} \sigma_2^S \otimes \tau^{A_4} },
    \end{align}
where:
	\begin{align}
\tilde{\tau}^{A_3}(m; \rho_0^S, \sigma_1^S, \tau^{A_2}, \tau^{A_3}) := \ptr{S A_2}{ \Pi_2^{(m)} \left( \mathcal{U}_{2:1}^{SA_2} \tilde{\tau}^{A_2}(\rho_0^S, \tau^{A_2}) \otimes \mathcal{U}_{2:1}^{SA_3} \sigma_0^S \otimes \tau^{A_3} \right) \Pi_2^{(m)}}.
	\end{align}
Without knowledge of $\rho_0^S$, the output state $\rho_3^{\prime S}$ when this instrument sequence is applied cannot be specified and hence the process displays memory effects that persist longer than $\ell$ timesteps.

\subsection{Other Generalized Collision Models with Memory}\label{app:othercm}

\begin{figure}[t]
\centering
\includegraphics[width=\linewidth]{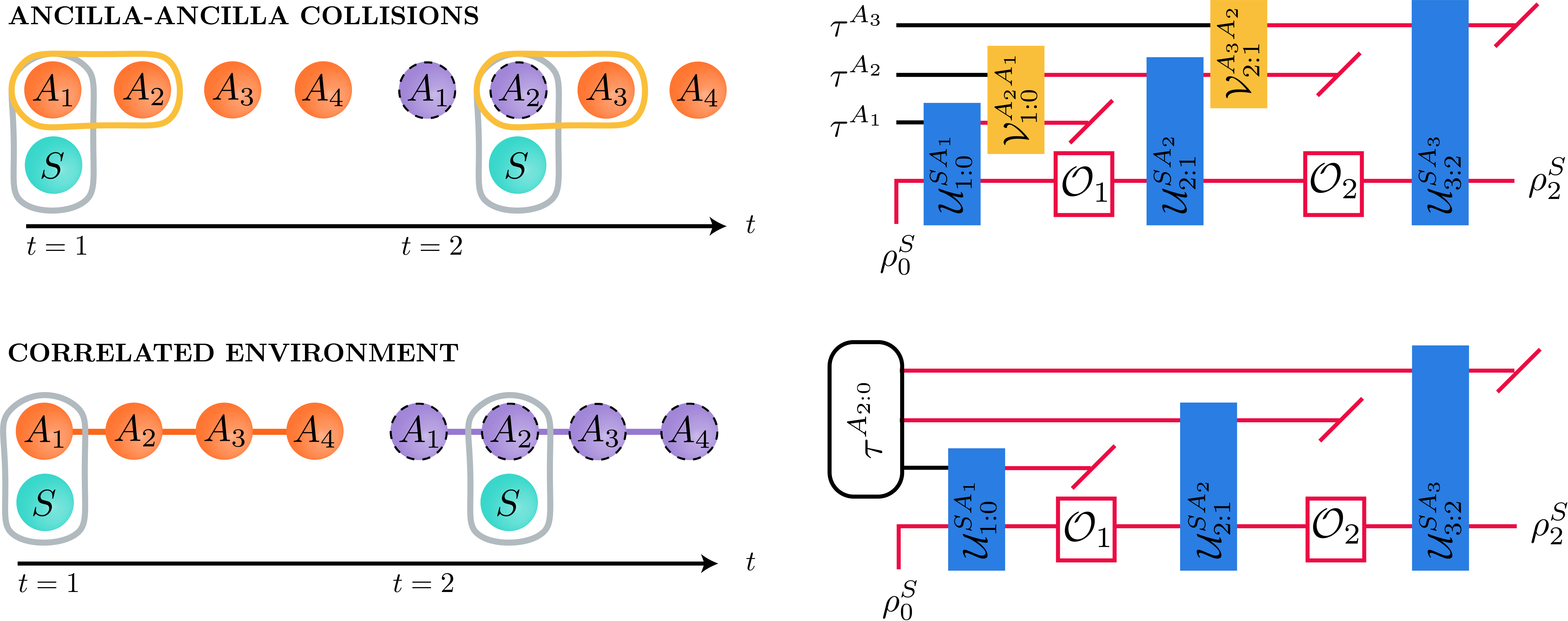}
\caption{\emph{Generalized collision models with memory.} In addition to introducing memory into collision models with repeated system-ancilla collisions (as in Section~\ref{sec:collisionmodel}), memory can be built into collision models by allowing for: \textbf{i)} ancilla-ancilla interactions (top row) and \textbf{ii)} an initially correlated environment (bottom row). (Note that the legend is as per Figs.~\ref{fig:collisionmodel} and~\ref{fig:trashingdilation}). The top left panel depicts a schematic of the dynamics where ancilla-ancilla interactions (yellow boundary) are interleaved between the system-ancilla collisions (gray boundary). At the second timestep, $A_2$ already has knowledge of the state of the system at $t=1$ mediated via the $A_1 A_2$ interaction. As such, the future dynamics is conditioned on the initial system state. The top right panel displays the corresponding circuit diagram. Here, for arbitrary operations on the system $\mathcal{O}_j$, it is clear that the ancilla-ancilla interactions provide a possible path of influence from the history to the future state; hence, such a process generically displays infinite Markov order with respect to any instrument sequence (as shown by the red path). For instance, even with the application of a sequence of trash-and-prepare operations, the final output state can still be influenced by the initial preparation. The bottom left panel depicts a schematic of the dynamics where the ancillas constituting the environment begin in a correlated state (represented by the orange line connecting them). Here, as soon as the system interacts with a part of the correlated ancillary state, all other ancillary systems can store information about the initial state of the system, thereby influencing the future dynamics. Finally, the bottom right panel displays the corresponding circuit diagram for this case. Again, allowing for arbitrary operations on the system level, the initial correlations in the environment provide a mechanism for the history to influence the future over an infinite length of time. \label{fig:othercm}}
\end{figure}

The example introduced in Section~\ref{sec:collisionmodel} presents a generalization of a collision model with memory; however, it is not the only way to build memory into collision models, which we now briefly explore. A discrete-time, $n$-step memoryless collision model consists of a system $S$ interacting with an environment $E$ that has a particular structure: it is made up of a number of constituent ancillary subsystems, $A_j$, with the dynamics proceeding through successive unitary collisions between the system and ancillas (see the top panel in Fig.~\ref{fig:collisionmodel}). A memoryless collision model assumes the following: 
\begin{enumerate}
    \item The system only interacts with each ancilla once.
    \item There are no ancilla-ancilla interactions.
    \item The ancillas are initially uncorrelated.
\end{enumerate}
Such a model has surprising power in describing dynamics which, in the continuous-time limit, are governed by a Lindbladian master equation~\cite{Brun2002,Ziman2005}. Although any such process looks Markovian at the system level, the necessary and sufficient Markov condition introduced in Ref.~\cite{Pollock2018L} only deems a process to be Markovian if the underlying $SE$ dynamics is exactly as described above. Breaking any one of the above assumptions (while satisfying the other two) endows the process with a different type of memory mechanism~\cite{Ciccarello2017} (see Figs.~\ref{fig:collisionmodel},~\ref{fig:trashingdilation}, and~\ref{fig:othercm} for illustration). We now examine such memory effects in terms of the structure of the underlying dilation, without any assumptions on the \emph{action} of the unitaries.


\textbf{Case 1: Repeated System-Ancilla Interactions.---}As shown in Section~\ref{sec:collisionmodel} and Appendix~\ref{app:cm}, in the case where one allows for repeated system-ancilla interactions, as in Refs.~\cite{Grimsmo2015,Whalen2017}, the memory effect depends on the nature of these repeated collisions. For example, if they occur in the nested order depicted in Fig.~\ref{fig:trashingdilation}, then the process has Markov order $\ell$ with respect to the trash-and-prepare protocol. If the interactions are simply repeated between the system and a given ancilla multiple times between each timestep, then the process is Markovian on an appropriate coarse-graining of timesteps. In general, however, repeated system-ancilla interactions give rise to infinite-length memory (even with respect to the trash-and-prepare protocol). This can be seen by considering the dynamics depicted in Fig.~\ref{fig:trashingdilation} with the order of any pair of joint unitary operations flipped: now, a continuous path can be drawn from the history to the future across a length-$\ell$ trash-and-prepare protocol, indicating a possible historic influence on the future dynamics.

\textbf{Case 2: Ancilla-Ancilla Interactions.---}This includes the scenarios considered in Refs.~\cite{Ciccarello2013,McCloskey2014,Lorenzo2016,Cakmak2017,Campbell2018} and is depicted in the top row of Fig.~\ref{fig:othercm}. In the case where ancilla-ancilla interactions are allowed, the historic influence can, in principle, last forever, since it can permeate continuously through the environment by ancilla-ancilla interactions. Consider specifically the case where at the first step, $S$ is swapped with $A_1$ through the swap map $\mathcal{U}_{1:0}^{SA_1} = \mathcal{S}^{SA_1}$, and then during each successive ancilla-ancilla interaction, the initial system state is continually swapped into the next ancilla via $\mathcal{S}^{A_j A_{j-1}}$, before finally $A_{n}$, which now stores the initial system state, is swapped back to the system level through $\mathcal{U}_{n:n-1}^{SA_n} = \mathcal{S}^{SA_n}$. Suppose that all but the first and last system-ancilla interactions are identity transformations and we allow for the application of arbitrary probing operations on the system at each timestep in between (represented by $\mathcal{O}_j$, which could, \textit{e.g.}, be trash-and-prepare operations). It is clear that the output system is (trivially) a function of its initial state, regardless of whatever intermediary operations an experimenter applies on the system:
\begin{align}
	\rho_n^S &= \ptr{A_n}{\mathcal{S}^{S A_n} \mathcal{O}_{n-1}^S \ptr{A_{n-1}}{ \mathcal{S}^{A_n A_{n-1}} \mathcal{O}_{n-2}^S \ptr{A_{n-2}}{\hdots \ptr{A_2}{\mathcal{S}^{A_3 A_2} \mathcal{O}_1^S \ptr{A_1}{\mathcal{S}^{A_2 A_1} \mathcal{S}^{S A_1}\rho_0^S \otimes \tau^{A_1} \otimes \hdots \otimes \tau^{A_n} }}}}} \notag \\
&= \ptr{A_n : A_1}{\mathcal{S}^{S A_n}  \mathcal{S}^{A_n A_{n-1}} \hdots \mathcal{S}^{A_{2} A_{1}} \mathcal{O}_{n-1}^S \hdots \mathcal{O}_{1}^S \mathcal{S}^{S A_1}\rho_0^S \otimes \tau^{A_1} \otimes \hdots \otimes \tau^{A_n} } \\
&= \ptr{A_n}{\mathcal{S}^{S A_n} \rho_0^{A_n} \mathcal{O}^S_{n-1:1} \tilde{\tau}^S } = \rho_0^S.  \notag
\end{align}
Here, we made use of the composition property of the swap map $\mathcal{S}^{A B} \mathcal{S}^{B C} = \mathcal{S}^{A C}$, defined $\mathcal{O}^S_{n-1:1} := \mathcal{O}^S_{n-1} \hdots \mathcal{O}^S_{1}$, and $\tilde{\tau}^S := \ptr{A_1}{\mathcal{S}^{SA_1} \rho_0^S \otimes \tau^{A_1}}$ is the initial state of $A_1$ that is swapped into the system space during the first joint interaction.

Despite the generally infinite-length memory, from the perspective of simulation, this type of memory is not complex: here, given control over part of the environment, one only needs to track one additional ancilla to efficiently simulate such processes, hence the classification of a ``memory depth'' of 1~\cite{Campbell2018}, even though the memory length here is infinite. Memory depth is the number of additional ancillas required to embed a non-Markovian process as a Markovian one; in other words, a process with a single ancilla-ancilla interaction between timesteps evolves in a Markovian fashion with respect to treating the system and the ancilla it interacts with at each timestep as a single larger system. In distinction, memory length concerns the number of timesteps back one needs to store information about the state of the system that could influence future dynamics. 

\textbf{Case 3: Initially Correlated Environment.---}Lastly, consider the case of an initially correlated environment, as is studied in Refs.~\cite{Rybar2012,Bernardes2014} and is depicted in the bottom row of Fig.~\ref{fig:othercm}. Again, there is no generic way to erase the influence of the state's history on its future evolution by action on the system alone: this is because the ancillary states in the environment begin correlated, and so as soon as the system interacts with the first ancilla, in principle all of the ancillas that \emph{will} interact with the system at some time in the future already store knowledge of the initial system state. Thus, through later interactions, this information can feed back to dictate the future evolution of the system, giving rise to non-Markovian behavior.

In the case of an initially correlated environment, one requires control over the entire collection of ancillas to simulate general processes. Again, consider the situation where $A_1$ and $A_{n-1}$ begin correlated, and at the first interaction $S$ and $A_1$ are swapped, and due to the $A_1$--$A_{n-1}$ correlation $A_{n-1}$ also stores knowledge of the initial system state, which can be swapped back to the system level at the final interaction to give the final output. At all intermediate timesteps, the dynamics looks like the initial state of $A_1$ interacting with each other ancilla pairwise in succession. It is clear that, as in case 2, the final state of the system will be identical to its initial state, regardless of the operations one might perform. However, in contrast, simulation of such processes is generically highly complex, as it requires control over a large number of ancillary subsystems in the environment.


\section{Choi-Jamio{\l}kowski Isomorphism for the Process Tensor}\label{app:cji}

\begin{figure*}[t]
\centering
\includegraphics[width=0.85\linewidth]{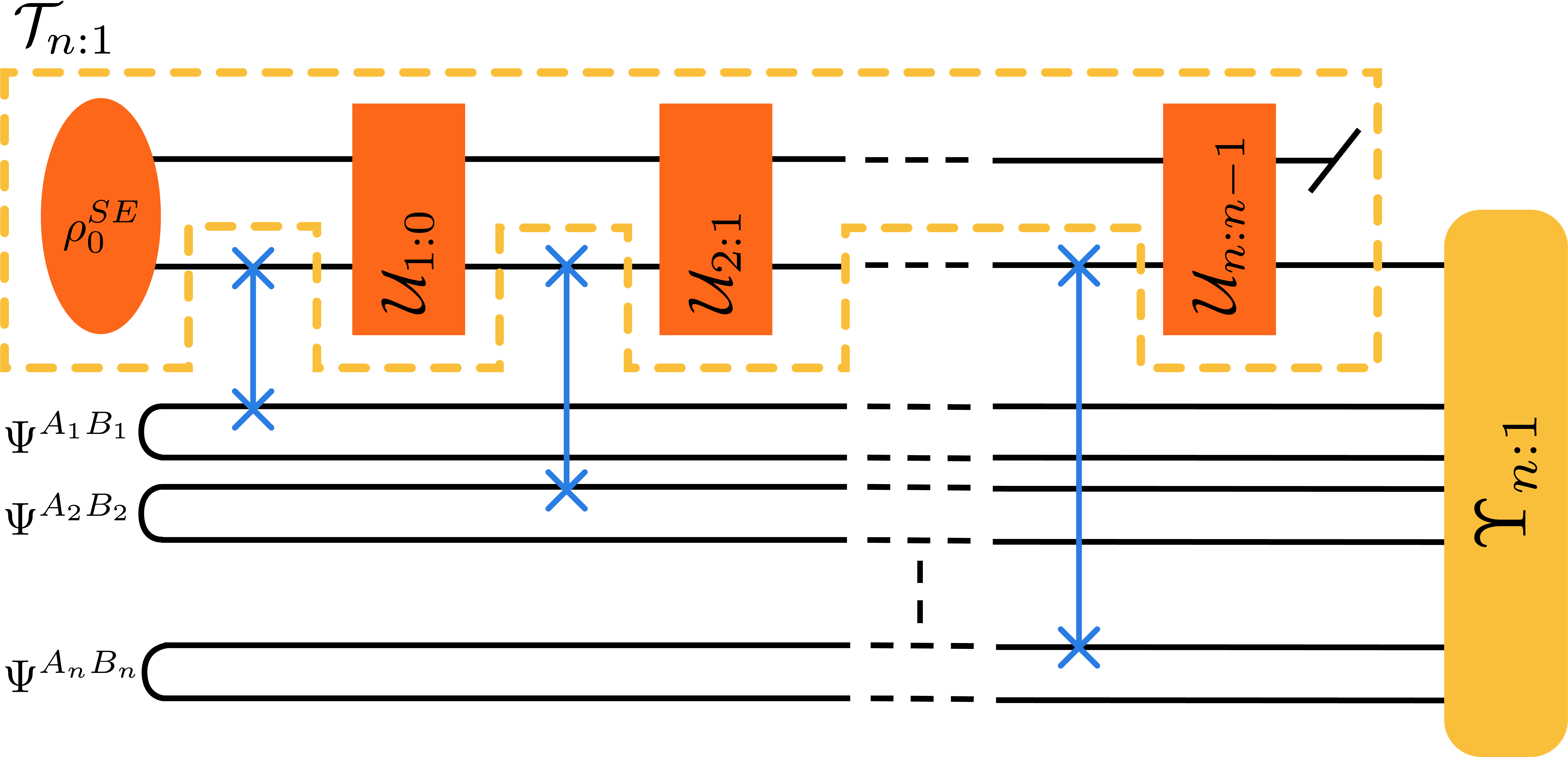}
\caption{\emph{Choi-Jamio{\l}kowski isomorphism for the process tensor.} The process tensor map, $\mathcal{T}_{n:1}$, for any quantum stochastic process can be represented as a many-body state, $\Upsilon_{n:1}$, through the extended Choi-Jamio{\l}kowski isomorphism depicted above and as described in Eq.~\eqref{eq:cji}. For each timestep $j \in \{ 1, \hdots, n \}$, half of an ancillary system that has been prepared as an unnormalized maximally entangled state, $\Psi^{A_j B_j}$, is swapped into the process, represented by the blue crosses. The resulting $(2n-1)$-body quantum state $\Upsilon_{n:1}$ contains equivalent information to the map $\mathcal{T}_{n:1}$. \label{fig:ptcji}}
\end{figure*}

In this appendix, we explicitly construct the Choi state of the process tensor, which is depicted in Fig.~\ref{fig:ptcji} and achieved as follows. Referring to Eq.~\eqref{eq:processtensoraction}, we abstract everything in the dynamics that is \emph{uncontrollable} as a multi-linear map, $\mathcal{T}_{n:1}$, which takes the \emph{control} operations implemented to the final state of the system, \emph{i.e.}, $\rho_{n}^S := \mathcal{T}_{n:1}[\mathcal{O}_{n-1}, \hdots, \mathcal{O}_{1}]$. Begin with $2n$ ancillary systems $\{A_j, B_j\}$ of the same dimensionality as $S$ prepared as $n$ (unnormalized) maximally entangled pairs, $\{\Psi^{A_j B_j}\}$. At each timestep of the process, one half of each pair is swapped with the system state through $\mathcal{S}^{S A_j}_j$. The resultant $d^{2n-1}$ dimensional system-ancillary (unnormalized) state $\Upsilon_{n:1} \in \mathcal{B}(\mathcal{H}^S \bigotimes_{j=1}^{n-1} \mathcal{H}^{A_j B_j})$ encodes equivalent information as the temporal map $\mathcal{T}_{n:1}$ and is explicitly written as follows: 
\begin{align}\label{eq:cji}
    \Upsilon_{n:1} &= \ptr{E}{\mathcal{U}_{n:n-1} \mathcal{S}^{S A_{n-1}}_{n-1} \hdots \mathcal{U}_{2:1} \mathcal{S}^{S A_{1}}_{1} \left(\rho^{SE}_0 \otimes \Psi^{A_1 B_1} \otimes \hdots \otimes \Psi^{A_n B_n} \right) } = \mathcal{T}_{n:1} \otimes \mathcal{L} \left[\mathcal{S}^{S A} \otimes {\mathcal{I}^{B}}\right],
\end{align}
where in the first equality, the $\mathcal{U}_{j:j-1}$ maps act only on the $SE$ space. The second equality makes the connection to the process tensor map $\mathcal{T}_{n:1}$ explicit; its action is extended to the ancillary space via a number of superchannels, $\mathcal{L} := \bigotimes_{j=1}^{n-1} \mathcal{L}^{A_j B_j}$, describing the initial maximally entangled pairs $\Psi^{AB}$, with $A$ accessible at timestep $j$, before both being subject to trivial evolution. This entire mapping acts on a number of swap operations $\mathcal{S}^{S A} := \bigotimes_{j=1}^{n-1} {\mathcal{S}^{S A_j}_j}$ between the system and the $A$ ancillas at each timestep, in addition to a trivial evolution of all of the $B$ ancillas, $\mathcal{I}^{B}$. 


Although any temporal process can be represented by a many-body quantum state through the above construction, as mentioned in the main text, not all quantum states correspond to a process. Eq.~\eqref{eq:causalconstraintmain} encodes the causality constraint required to ensure that the Choi state represents a valid temporal process. Since the process tensor acts on sequences of CP maps, it makes no sense \emph{per se} to speak of the standard desired properties of a valid process such as complete-positivity and trace-preservation; on the other hand, there are natural extensions of these concepts that must be satisfied~\cite{Chiribella2008,Pollock2018A}. The notion of complete-positivity means that for any sequences of input CP maps, including those acting on an arbitrarily extended ancillary Hilbert space, the output of the process tensor is a valid CP map: this leads to the positivity and Hermiticity of the Choi state $\Upsilon_{n:1} \geq 0$ and $\Upsilon_{n:1}^\dagger = \Upsilon_{n:1}$. The trace-preservation property translates to the statement that for any deterministic sequences of operations applied \textit{i.e.} CPTP maps, the output state has unit trace: this is encoded naturally in Eq.~\eqref{eq:causalconstraintmain} by the fact that the partial trace over the final system yields an identity operator on the output space of the previous timestep (rather than a subnormalized state proportional to the identity). Indeed, positive, Hermitian operators satisfying Eq.~\eqref{eq:causalconstraintmain} are the most general (unnormalized) quantum states that represent physically allowable evolutions through the (inverse of the) Choi-Jamio{\l}kowski isomorphism defined in Eq.~\eqref{eq:cji}. 

The action of the process tensor map on a sequence of operations can be expressed in terms of its Choi state (and that of the sequence of operations it acts upon)
\begin{align}
    \mathcal{T}_{n:1}[O_{n-1:1}] = \ptr{n-1:1}{(\mathbbm{1}_n^\inp \otimes O_{n-1:1}^\text{T}) \Upsilon_{n:1}},
\end{align}
where $O_{n-1:1}$ represents the most general correlated sequence of CP maps an experimenter could apply. Indeed, any such sequence of CP maps must be physically implementable, and so are themselves similar objects to the process tensor and subject to a complementary set of causality constraints. Generalizing the notion of an instrument, an \emph{instrument sequence} is any such collection of (possibly correlated across timesteps) CP maps $\mathcal{J}_{n-1:1} := \{ O_{n-1:1}^{(x_{n-1:1})} \}$ that, overall, give rise to a valid process, \textit{i.e.}, $\widetilde{O}_{n-1:1} := \sum_{x_{n-1:1}} O_{n-1:1}^{(x_{n-1:1})}$ with $\widetilde{O}_{n-1:1}$ being a valid quantum state satisfying Eq.~\eqref{eq:causalconstraintmain}. 

Lastly, we present a summary of the notation consistently used throughout this Article to aid the reader. Timesteps are labeled as subscripts with Latin letters; to ease notation, sequences are often grouped as $j:k := \{ j , \hdots, k\}$. The labels $\inp / \out$ refer to the input (respectively output) Hilbert spaces associated to each timestep (from the perspective of somebody applying operations to the system state). Uppercase script Latin letters denote temporal maps and their non-script counterparts represent their corresponding Choi state. Superscripts in parentheses label outcomes of probing instruments. Capitalized Latin superscripts are used to denote spaces of operators where they may be potentially ambiguous. 

\section{Vanishing Quantum CMI implies Finite Quantum Markov Order} 
\label{app:proofcmi}

\begin{proof}\textbf{(Theorem~\ref{thm:cmi}).} From the structure of Eq.~\eqref{eq:qcmistates}, it is clear that there exists a history-blocking instrument sequence: namely, that constituting the projectors onto each of the $m$ orthogonal subspaces. Begin by rewriting Eq.~\eqref{eq:qcmistates} as a regular sum by projecting onto the constituent subspaces of the decomposition:
\begin{align}\label{eq:cmiproof1}
    \Upsilon_{FMH}^{CMI=0} &= \bigoplus_{m} \mathbbm{P}(m) \Upsilon_{F M^L}^{(m)} \otimes \Upsilon_{M^R H}^{(m)} \\
    &= \sum_m \mathbbm{P}(m) \Pi_{M^L}^{(m)} \Upsilon_{F M^L}^{(m)} \Pi_{M^L}^{(m)} \otimes \Pi_{M^R}^{(m)} \Upsilon_{M^R H}^{(m)}  \Pi_{M^R}^{(m)}. \notag
\end{align}
Consider now the instrument made up of the projectors in the above decomposition, \emph{i.e.}, $\mathcal{J}_M = \{ \Pi_{M^L}^{(m)} \otimes \Pi_{M^R}^{(m)} \}$. This constitutes a valid instrument sequence as it sums to an identity on $\mathcal{H}_M$. It also constitutes a history-blocking sequence for the process described by $ \Upsilon_{FMH}^{CMI=0}$, as for each realization of the instrument, the future and history are conditionally independent:
\begin{align}\label{eq:cmiproof2}
    \ptr{M}{\left(\Pi_{M^L}^{(m^\prime)} \otimes \Pi_{M^R}^{(m^\prime)}\right) \Upsilon_{FMH}^{CMI=0}} &= \ptr{M}{\sum_m \mathbbm{P}(m) \Pi_{M^L}^{(m)} \Upsilon_{F M^L}^{(m)} \Pi_{M^L}^{(m)} \otimes \Pi_{M^R}^{(m)} \Upsilon_{M^R H}^{(m)}  \Pi_{M^R}^{(m)} \left( \Pi_{M^L}^{(m^\prime)} \otimes \Pi_{M^R}^{(m^\prime)} \right)} \\
    &= \ptr{M}{ \Upsilon_{F M^L}^{(m^\prime)} \Pi_{M^L}^{(m^\prime)} \otimes \Upsilon_{M^R H}^{(m^\prime)} \Pi_{M^R}^{(m^\prime)}} \notag \\
    &= \ptr{M^L}{ \Upsilon_{F M^L}^{(m^\prime)}} \otimes \ptr{M^R}{ \Upsilon_{M^R H}^{(m^\prime)} } \notag \\
    &= \Upsilon_{F}^{(m^\prime)} \otimes \Upsilon_{H}^{(m^\prime)}, \notag
\end{align}
where we use the orthogonal projector identity $\Pi^{(m)} \Pi^{(m^\prime)} = \delta_{m m^\prime } \Pi^{(m)}$ and the trace properties of cyclicity and linearity. 

We now examine the structure of vanishing quantum CMI processes in further detail: this serves to illuminate the connection between processes with finite Markov order with respect to instruments comprising only orthogonal projectors and those with vanishing quantum CMI, which we explore further in Section~\ref{sec:vanishingqcmi}. Continuing from Eq.~\eqref{eq:cmiproof1}, note that the projectors in the decomposition are not necessarily rank-1; we thus further expand the conditional process tensor parts on a basis within each $m$ subspace as
\begin{align}\label{eq:cmiproof3}
    \Upsilon_{F M^L}^{(m)} \otimes \Upsilon_{M^R H}^{(m)} = \Upsilon_F^{(m)} \otimes \Pi_{M^L}^{(m)} \otimes \Pi_{M^R}^{(m)} \otimes \Upsilon_H^{(m)} + \sum_{s s'} \Upsilon_F^{(m,s)} \otimes \xi_{M^L}^{(s)} \otimes \xi_{M^R}^{(s^\prime)} \otimes \Upsilon_H^{(m,s')}.
\end{align}
The $\xi_{M^{L/R}}$ encode the off-diagonal elements within each $m$ subspace (since the projector $\Pi_{M^L}^{(m)} \otimes \Pi_{M^R}^{(m)}$ encodes all of the diagonal elements), and can thus be chosen such that $\tr{\xi_{M^{L/R}}} = 0$ and $\Pi_{M^{L/R}}^{(m)} \xi_{M^{L/R}}^{(m^\prime)} = \delta_{m m^\prime} \xi^{(m)}_{M^{L/R}}$. In this expansion, the $\Upsilon_F^{(m,s)}$ and $\Upsilon_H^{(m , s^\prime)}$ are not required to be proper processes, since the $\xi_{M^{L/R}}$ do not necessarily represent physical operators. We therefore have
\begin{align}\label{eq:cmiproof4}
    &\Upsilon_{FMH}^{CMI=0} = \sum_m \mathbbm{P}(m) \Upsilon_F^{(m)} \otimes \Pi_{M^L}^{(m)} \otimes \Pi_{M^R}^{(m)} \otimes \Upsilon_H^{(m)} + \sum_{m, s, s'} \Upsilon_F^{(m, s)} \otimes \xi_{M^L}^{(m, s)} \otimes \xi_{M^R}^{(m, s^\prime)} \otimes \Upsilon_{H}^{(m, s^\prime)}. 
\end{align}
Note that if the $M$ subspaces in the decomposition of Eq.~\eqref{eq:qcmistates} are all one dimensional, \emph{i.e.}, the projectors in Eq.~\eqref{eq:cmiproof1} are all rank-1, we only have the first term:
\begin{align}\label{eq:cmiproof5}
    &\Upsilon_{FMH}^{CMI=0} = \sum_m \mathbbm{P}(m) \Upsilon_F^{(m)} \otimes \Pi_{M^L}^{(m)} \otimes \Pi_{M^R}^{(m)} \otimes \Upsilon_H^{(m)}.
\end{align}

Regarding the converse statement of Theorem~\ref{thm:cmi}, we have shown examples of processes with finite Markov order with non-vanishing quantum CMI (see Examples~\ref{ex:unitary} and \ref{ex:causalbreak} and the generalized collision model of Section~\ref{sec:collisionmodel}); processes with finite Markov order must only satisfy the structure outlined in Theorem~\ref{thm:markovorderstructure}. This shows that it is insufficient to conclude that the process has vanishing quantum CMI. Furthermore, even if a given process has finite Markov order with respect to an instrument sequence comprising only rank-1, orthogonal projectors, the process can still have non-vanishing quantum CMI: in this case, since the projectors form a self-dual set, we can construct the process as
\begin{align}\label{eq:cmiproof6}
    \Upsilon_{FMH} =& \sum_{m} \Upsilon_F^{(m)} \otimes \Pi_M^{(m)} \otimes \Upsilon_H^{(m)} + \widetilde{\Upsilon}_{FMH}, 
\end{align}
with $\tr{   \Pi_M^{(m)} \widetilde{\Upsilon}_{M}} = 0 \, \forall \, m$. Even though the projectors in the history-blocking instrument are not necessarily the same as those that project onto the subspaces defined in the decomposition of Eq.~\eqref{eq:qcmistates}, this condition does not imply that the process tensor is block-diagonal in some basis of $M$; rather, the process can have off-diagonal elements with respect to the subspaces defined by $\{ \Pi_M^{(m)}\}$ and satisfy Eq.~\eqref{eq:cmiproof6}. This implies that there are processes with non-vanishing quantum CMI but finite Markov order.

\end{proof}

To summarize, the salient points from this analysis are as follows. Firstly, suppose that a process has finite Markov order with respect to an instrument sequence comprising only orthogonal projectors that are not rank-1: in this case, there is no reason that the future-history correlations within each $m$ subspace must obey the product structure outlined in Eq.~\eqref{eq:cmiproof5}, and hence the process can have non-vanishing quantum CMI. This is shown explicitly in Example~\ref{ex:nonsharpproj} of Appendix~\ref{app:quantumfuzzy}. However, similar behavior also arises in an operational interpretation of classical stochastic processes, as explored in Section~\ref{sec:classical}: if one cannot measure realizations of the process \emph{sharply},~\emph{i.e.}, with sequences of rank-1 projectors, then the statistics observed do not necessarily have vanishing classical CMI, even if the true underlying process is one of finite Markov order (see Appendix~\ref{app:classical}). 

Secondly, suppose that a process has finite Markov order with respect to an instrument sequence comprising sharp, orthogonal projectors. The condition $\tr{  \Pi_M^{(m)} \widetilde{\Upsilon}_{M}  } = 0 \, \forall \, m$ of Eq.~\eqref{eq:cmiproof6} does not imply that the process must be block-diagonal in some basis of $\mathcal{H}_M$, as required for the quantum CMI to vanish [see Eq.~\eqref{eq:cmiproof1}]. It follows that there exist such processes with non-vanishing quantum CMI, as shown in Example~\ref{ex:qcmi}. In contrast to the earlier point regarding instrument sequences comprising higher-rank projectors, the present statement is indeed a fundamentally quantum mechanical phenomenon. In the classical setting, finite Markov order with respect to sharp realizations of the process and the classical CMI vanishing are equivalent statements (see Section~\ref{sec:classical}). 

It is lastly interesting to consider why these two notions are equivalent in the classical setting but not for quantum processes. Suppose that a classical process has finite Markov order with respect to the sequence of sharp projectors $\{ \Pi_M^{(m)} \}$; then, the process can be written of the form in Eq.~\eqref{eq:cmiproof6}. However, in the classical setting, where there can be no off-diagonal terms, $\tr{  \Pi_M^{(m)} \widetilde{\Upsilon}_{M} } = 0$ implies $\widetilde{\Upsilon}_M = 0$. Alternatively, $d$ orthogonal projectors are informationally-complete in the classical setting; thus, the process must be of the form of Corollary~\ref{cor:complete}, with the projectors on the $M$ block. In either case, the process is then of the form of Eq.~\eqref{eq:cmiproof5} (by choosing either $\mathcal{H}_{M^L}$ or $\mathcal{H}_{M^R}$ to be trivial), meaning the quantum CMI vanishes. 

\section{Classical Stochastic Processes with Fuzzy Measurements}\label{app:classical}

Here we provide two examples, depicted in Fig.~\ref{fig:coarsegrain}, of classical processes where an experimenter has only access to a fuzzy measuring device which coarse-grains over some of the outcomes of the process at hand: in either case, the perceived memory length of the process is instrument-dependent. The first example is a process that is Markovian but exhibits non-Markovian statistics to the experimenter, while the second example is a process that is non-Markovian but looks Markovian on average, \emph{i.e.}, with respect to the coarse-graining instrument.

\begin{example}\label{ex:classical1} \textit{Fuzzy Measurements can Increase Classical Markov Order}.---Consider the process depicted in the left panel of Fig.~\ref{fig:coarsegrain}. At each timestep, the random variable of interest $X_k$ can take one of three distinct values: $x_k \in \{ a_k, b_k, c_k\}$. Between each step of dynamics, the transition probabilities are given by $\mathbbm{P}_k(b_k|a_{k-1}) = \mathbbm{P}_k(c_k|b_{k-1}) = 1$, $\mathbbm{P}_k(a_k|c_{k-1}) = p$, $\mathbbm{P}_k(b_k|c_{k-1}) = 1-p$ [with $p \in (0,1)$] and all other transitions are forbidden. Such a process is clearly Markovian in the random variable $X$, as knowledge of any current state suffices to deduce the probability of the next. Now suppose that for the same process, one could not distinguish between outcomes $b_k$ and $c_k$ at each timestep: \textit{i.e.}, instead of $X$, we observe the random variable $Y$, which takes values $y_k \in \{ a_k, d_k =  b_k \cup c_k \}$. In this case, when the state at some time is $a$, the next state is for sure $d$; while if the state is $d$, with probability $p$ it will transition next to $a$ or with probability $1-p$ it will remain $d$ (alternating between $b$ and $c$ deterministically, although we are ignorant of this fact). Thus, conditioned on any consecutive $j$ observations of outcome $d$ following an observation of $a$:
\begin{align}
    \mathbbm{P}_k(a_k|d_{k-1}, \hdots, d_{k-j}, a_{k-j-1}) = \begin{cases}
    0 \quad \text{j odd} \\
    p \quad \text{j even.}
    \end{cases}
\end{align}
With this fuzzy measurement apparatus at hand, one would consider the process to be non-Markovian with respect to realizations of $Y$. Lastly, note that given an instrument that alternatively measures $X$ and then $Y$ at each consecutive timestep, the experimenter would determine the Markov order to be $\ell=2$. 


\begin{figure}[t]
\centering
\includegraphics[width=0.88\linewidth]{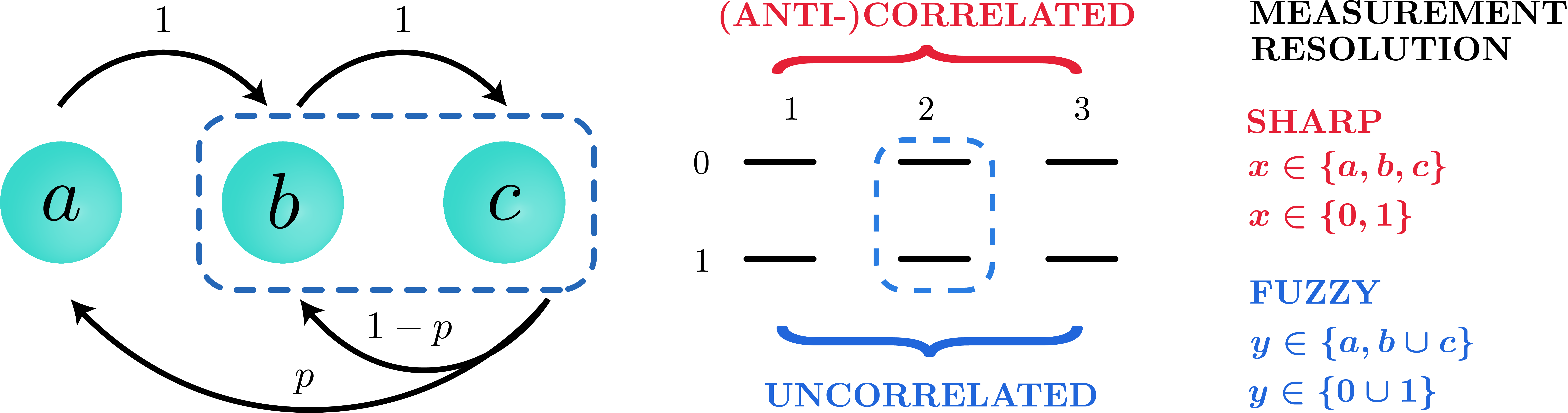}
\caption{\emph{Instrument-dependence of classical Markov order with fuzzy measurements.} Here we depict two classical processes to highlight the instrument-dependence of Markov order when sharp measurements are not assumed (a legend is provided in the rightmost panel). On the left, the process of Example~\ref{ex:classical1} is shown, defined by the transition probabilities depicted at each timestep. Here, if one is able to record observations sharply, \textit{i.e.}, measure the random variable $X = \{ a, b, c\}$, the process is clearly Markovian; however, if one cannot measure at that resolution and, \textit{e.g.}, the measurement apparatus only records fuzzy statistics of the random variable $Y = \{ a,  b \cup c \}$, as depicted by the blue, dashed box, the process would be classified as non-Markovian. In the middle panel, the process of Example~\ref{ex:classical2} is shown. Here, three bits are initially prepared as described in the text, and each bit is fed out of the process at successive timesteps. The preparation is such that if the second bit is sharply measured to be in the state 0, bits 1 and 3 are perfectly correlated; if the second bit is in state 1, bits 1 and 3 are perfectly anti-correlated; while on average, \emph{i.e.}, with respect to the fuzzy measurement coarse-graining over everything in the blue, dashed box, bits 1 and 3 are completely uncorrelated. \label{fig:coarsegrain}}
\end{figure}


\end{example}

\begin{example}\label{ex:classical2} \textit{Fuzzy Measurements can Decrease Classical Markov Order}.---Consider the process depicted in the right panel of Fig.~\ref{fig:coarsegrain}. Here, three bits described by random variables $\{ X_1, X_2, X_3\}$ are fed out in succession over three timesteps. Suppose that these bits are initially prepared according to the probability distribution $\mathbbm{P}_{3:1}(X_3,X_2,X_1)$, which is such that $\mathbbm{P}_{3:1}(0,0,0) = \mathbbm{P}_{3:1}(1,0,1) = \tfrac{1}{4}$, $\mathbbm{P}_{3:1}(0,1,1) = \mathbbm{P}_{3:1}(1,1,0) = \tfrac{1}{4}$, and the rest of the terms vanish. The process is such that if the bit output at the second step is measured to be 0, then the first and third bits are perfectly (classically) correlated; while if bit at the second step is measured to be 1, then the first and third bits are perfectly (classically) anti-correlated. Thus, the process is perceived to be non-Markovian with respect to sharp measurements of the second bit value. On the other hand, on average, there is no correlation between the first and third bits; thus, with respect to a coarse-grained measurement of the second bit value, the process is perceived to be Markovian.
\end{example}

We conclude this appendix by explicitly phrasing these ideas in terms of the process tensor and instrument sequences that block the history. Firstly, any classical stochastic process can be encoded in process tensor that is diagonal in a fixed, local product basis of sharp, orthogonal projectors representing each outcome:
\begin{align}\label{eq:classicalPT}
    \Upsilon^{\text{Cl}}_{FMH} &= \sum_{x} \mathbbm{P}_{FMH}(x_F, x_M, x_H) \ket{x}\bra{x}_{FMH}^\inp \otimes \mathbbm{1}_{FMH}^\out,
\end{align}
where $\ket{x}\bra{x}_{FMH}^\inp = \ket{x}\bra{x}_{n^\inp} \otimes \hdots \otimes \ket{x}\bra{x}_{1^\inp}$. Suppose that we can measure sharply in the correct basis: this corresponds to applying projective operators of the form $P_k = \ket{x}\bra{x}_{k^\out} \otimes \ket{x}\bra{x}_{k^\inp}$ over a sequence of timesteps; overall, if we observe some realization of the process, we have implemented the instrument that is a collection of these operators. Indeed, the process tensor defined in Eq.~\eqref{eq:classicalPT} is diagonal with respect to the classical reference basis defined by $\{ \ket{x}_{FMH} \}$ and yields the correct statistics upon application of such sharp classical measurements. 

Now, suppose that the classical process at hand is one of finite Markov order; then, the joint probability distribution must satisfy Eq.~\eqref{eq:cmarkovorder}. It is straightforward to show that any length-$\ell$ instrument sequence comprising \emph{sharp}, orthogonal measurements $\{ P_{k-\ell}, \hdots, P_{k-1} \}$ provides a history-blocking sequence, rendering the history and future processes conditionally independent. On the other hand, \emph{fuzzy} classical measurements correspond to operators of the form $\ket{y}\bra{y}_{k^\out} \otimes \ket{x}\bra{x}_{k^\inp}$, where $\ket{y}\bra{y}_{k^\out} := \sum_{\overline{x}} \ket{x}\bra{x}_{k^\out}$ is a coarse-graining over some of the outcomes labeled by $\ket{x}\bra{x}_{k^\inp}$. In other words, although we \emph{measure} outcome $y$, the true state of the system that is fed forward into the process is, unbeknownst to us, $x$. In this case, the statistics observed would not satisfy Eq.~\eqref{eq:cmarkovorder}, but would rather be a mixture of statistics that do. Consequently, it would, in general, have a different Markov order than detected by the fine-grained measurements, and the classical CMI computed on the statistics observed would not vanish. 

\section{Quantum Analog of Example~\ref{ex:classical2}}\label{app:quantumfuzzy}

\begin{figure}[t]
\centering
\includegraphics[width=0.88\linewidth]{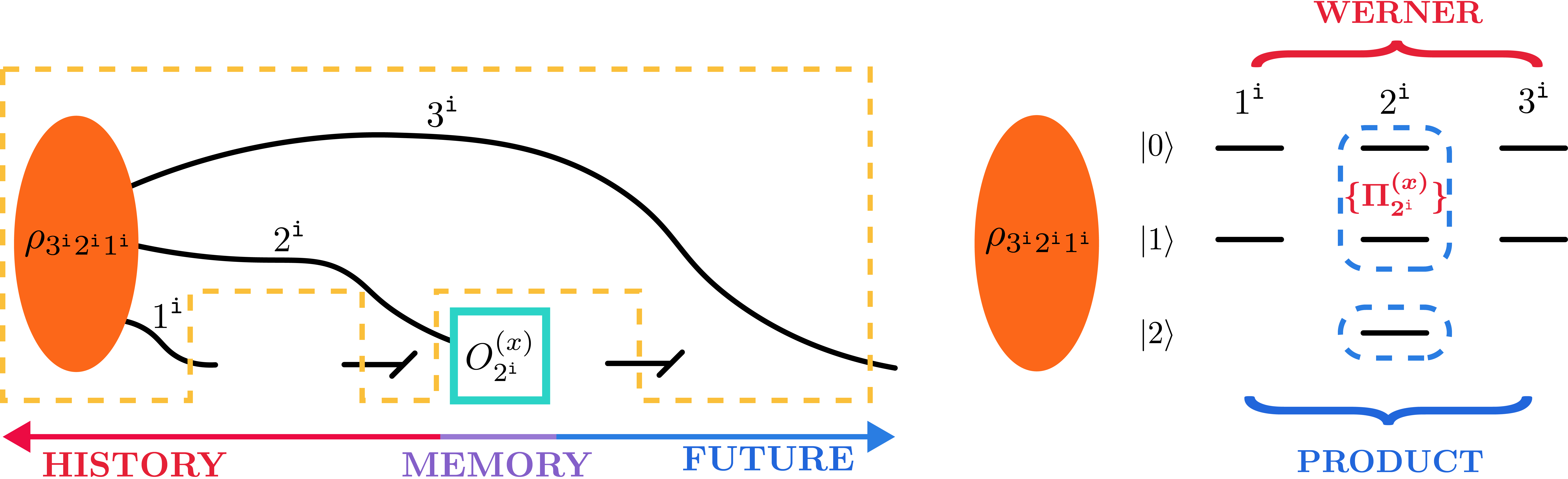}
\caption{\emph{Process with non-vanishing quantum conditional mutual information but finite Markov order with respect to (fuzzy) orthogonal projectors.} The tripartite state $\rho_{3^\inp 2^\inp 1^\inp}$ as defined in Eq.~\eqref{eq:totalcondwerner} is depicted on the left. Here, if an experimenter cannot distinguish between measurement outcomes in the $\{ \ket{0}, \ket{1}\}$ subspace of $\mathcal{H}_{2^\inp}$, represented by the blue, dashed boxes, then the conditional state $\rho_{3^\inp 1^\inp}^{(x)}$ for each outcome is product. If, on the other hand, the experimenter can resolve sharp measurements in the $\{ \ket{0}, \ket{1}\}$ subspace and implement, \textit{e.g.}, the operations $\{O^{(x)}_{2^\inp}\} = \{\Pi^{(x)}_{2^{\inp}}\}$, then for each outcome realized, the conditional state $\rho_{3^\inp 1^\inp}^{(x)}$ is a (correlated) Werner state, defined in Eq.~\eqref{eq:werner}. The process is such that this state is initially prepared, with subsystems fed out to the experimenter over a sequence of timesteps. Whatever is fed back into the process is discarded by the process itself; hence, the process tensor is written as per Eq.~\eqref{eq:ptcondwerner}. As described above, the fuzzy (orthogonal) measurement at timestep $2^\inp$ blocks the influence of history on the future, although a sharp measurement resolving all three outcomes does not. Lastly, the quantum CMI for this process does not vanish. \label{fig:nonsharpproj}}
\end{figure}

\begin{example}\label{ex:nonsharpproj}
\textit{Process with Non-Vanishing Quantum CMI but Finite Markov Order for a Sequence of Fuzzy, Orthogonal Projectors.}---Consider the process depicted in Fig.~\ref{fig:nonsharpproj}. Begin with the four two-qubit Werner states,
\begin{align}\label{eq:werner}
    \rho_{3^\inp 1^\inp}^{(x)}(r) := r \beta^{(x)} + (1-r) \frac{\mathbbm{1}}{2},
\end{align}
where $r \in (0, 1)$ and each $\beta^{(x)}$ is one of the four Bell pairs,
\begin{align}\label{eq:normalizedbell}
    \ket{\psi^\pm} := ( \ket{00}\pm\ket{11})/\sqrt{2} \quad \text{and} \quad
    \ket{\phi^\pm} := ( \ket{01}\pm\ket{10})/\sqrt{2}. 
\end{align}
Take some symmetric, IC qubit POVM $\{ \Pi^{(x)}_{2^{\inp}} \}$, and, in terms of its dual set $\{ \Delta^{(x)}_{2^{\inp}} \}$, construct the following state:
\begin{align}\label{eq:condwerner}
    \mu_{3^\inp 2^\inp 1^\inp }(r) := \sum_x \frac{1}{4} \rho_{3^\inp 1^\inp}^{(x)}(r) \otimes  \Delta^{(x)}_{2^{\inp}}.
\end{align}
This object is positive (and therefore a valid state) only for $r \in (0, 1/3]$, which corresponds to the values for which the Werner states defined in Eq.~\eqref{eq:werner} are separable. Now, let the system $\mathcal{H}_{2^\inp}$ represent a qutrit; the first two levels are described by Eq.~\eqref{eq:condwerner}, which is mixed with probability $q \in (0,1)$ with an arbitrary tensor product state $\rho_{3^\inp} \otimes \rho_{1^\inp}$ in product with the third-level basis state $\ket{2}$, giving the overall initial system-environment state:
\begin{align}\label{eq:totalcondwerner}
    \rho_{3^\inp 2^\inp 1^\inp }(q,r) &= q \mu_{3^\inp 2^\inp 1^\inp }(r) + (1-q) \rho_{3^\inp} \otimes \ket{2}\bra{2}_{2^\inp} \otimes \rho_{1^\inp}. 
\end{align}
The process proceeds by initially preparing this state and feeding out the $\rho_{j^\inp }$ marginal state at each timestep $j = \{ 1, 2, 3 \}$. No matter what operations are implemented on the system at these timesteps, the process acts to discard whatever is fed back into it; thereby, it has trivial output spaces and the corresponding process tensor is:
\begin{align}\label{eq:ptcondwerner}
    \Upsilon_{3^\inp :1^\inp}(q,r) = \rho_{3^\inp 2^\inp 1^\inp }(q,r) \otimes \mathbbm{1}_{2^\out 1^\out}.
\end{align}

Now, consider the instrument made up of the following two fuzzy, orthogonal operations $O^{(1)}_{2^\inp} = (\mathbbm{1} - \ket{2}\bra{2})_{2^\inp}$ and $O^{(2)}_{2^\inp} = \ket{2}\bra{2}_{2^\inp}$. With respect to this instrument, the conditional process tensors for each outcome are:
\begin{align}\label{eq:condptcondwerner}
    \Upsilon^{(1)}_{3^\inp 2^\out 1^\out 1^\inp } = \frac{\mathbbm{1}_{3^\inp}}{2} \otimes \mathbbm{1}_{2^\out 1^\out} \otimes \frac{\mathbbm{1}_{1^\inp}}{2} \quad \text{ and } \quad
    \Upsilon^{(2)}_{3^\inp 1^\out 1^\inp } = \rho_{3^\inp} \otimes \mathbbm{1}_{2^\out 1^\out} \otimes \rho_{1^\inp}. 
\end{align}
Thus, Eq.~\eqref{eq:qmarkovordercondition} is satisfied and the process has Markov order $1$ with respect to this instrument comprising only (fuzzy) orthogonal projectors (note that this process is not Markovian, as an IC instrument of causal breaks will not block the history). However, had the experimenter been able to resolve measurements in the $\{ \ket{0}, \ket{1} \}$ subspace of $\mathcal{H}_{2^\inp}$, \textit{e.g.}, apply the instrument comprising the operations $O^{(x)}_{2^\inp} = \Pi^{(x)}_{2^{\inp}}$ for $x \in \{ 1,2,3,4 \}$ and $O^{(5)}_{2^\inp} = \ket{2}\bra{2}_{2^\inp}$, then the conditional process tensors for each outcome are:
\begin{align}\label{eq:condptcondwernercorr}
    \Upsilon^{(x)}_{3^\inp 1^\out 1^\inp} = \psi^{(x)}_{3^\inp 1^\inp} \otimes \mathbbm{1}_{2^\out 1^\out} \quad \text{ and } \quad
    \Upsilon^{(5)}_{3^\inp 1^\out 1^\inp} = \rho_{3^\inp} \otimes \mathbbm{1}_{2^\out 1^\out} \otimes \rho_{1^\inp}.
\end{align}
For each outcome $x$ observed in the $\{ \ket{0}, \ket{1} \}$ subspace, the conditional future and history processes exhibit correlations via one of the four Werner states (which are separable, but not product, and thereby correlated). Similarly, if an experimenter applied the sharp constituent projectors that make up the fuzzy history-blocking instrument, \emph{i.e.}, measure the three outcomes associated to $\{ \ket{0}\bra{0}, \ket{1}\bra{1}, \ket{2}\bra{2}\}$, the conditional states for outcomes (0) and (1) are again correlated. Lastly, note that this process has non-vanishing quantum CMI: a straightforward calculation shows that $I(F:H|M) = q$ for $\Upsilon_{3^\inp :1^\inp}(q,r)$ defined in Eq.~\eqref{eq:ptcondwerner}. 

\end{example}


\newpage

\FloatBarrier

%

\end{document}